\documentclass[11pt,a4paper]{article}

\usepackage{jheppub}
\usepackage{amsfonts}
\usepackage{amssymb}
\usepackage{epsfig}
\usepackage{graphicx}
\usepackage{amsmath}
\usepackage{amsxtra}

\usepackage{color}

\def\circa#1{\,\raise.3ex\hbox{$#1$\kern-.75em\lower1ex\hbox{$\sim$}}\,}
\makeatletter
\def\art{\@ifnextchar[{\eart}{\oart}}
\def\eart[#1]#2#3#4#5#6{{\rm #2}, {\em #3  #4} {\rm (#6) #5} ({\em #1})}
\def\hepart[#1]#2{{\rm #2, \em#1}}
\newcommand{\oart}[5]{{\rm #1}, {\em #2  #3} {\rm (#5) #4}}

\newcounter{alphaequation}[equation]
\def\thealphaequation{\theequation\hbox to
0.6em{\hfil\alph{alphaequation}\hfil}}
\def\eqnsystem#1{
\def\@eqnnum{{\rm (\thealphaequation)}}
\def\@@eqncr{\let\@tempa\relax \ifcase\@eqcnt \def\@tempa{& & &} \or
\def\@tempa{& &}\or \def\@tempa{&}\fi\@tempa
\if@eqnsw\@eqnnum\refstepcounter{alphaequation}\fi
\global\@eqnswtrue\global\@eqcnt=0\cr}
\refstepcounter{equation} \let\@currentlabel\theequation \def\@tempb{#1}
\ifx\@tempb\empty\else\label{#1}\fi
\refstepcounter{alphaequation}
\let\@currentlabel\thealphaequation
\global\@eqnswtrue\global\@eqcnt=0 \tabskip\@centering\let\\=\@eqncr
$$\halign to \displaywidth\bgroup \@eqnsel\hskip\@centering
$\displaystyle\tabskip\z@{##}$&\global\@eqcnt\@ne
\hskip2\arraycolsep\hfil${##}$\hfil& \global\@eqcnt\tw@\hskip2\arraycolsep
$\displaystyle\tabskip\z@{##}$\hfil
\tabskip\@centering&\llap{##}\tabskip\z@\cr}
\def\endeqnsystem{\@@eqncr\egroup$$\global\@ignoretrue} \makeatother

\newcommand{\vo}{{\cal V}}

\newcommand{\roughly}[1]{\mathrel{\raise.3ex\hbox{$#1$\kern-0.85em
\lower1ex\hbox{$\sim$}}}}

\newcommand{\mc}{\mathcal}
\newcommand{\mbb}{\mathbb}
\newcommand{\bea}{\begin{eqnarray}}
\newcommand{\eea}{\end{eqnarray}}
\newcommand{\be}{\begin{equation}}
\newcommand{\ee}{\end{equation}}
\newcommand{\bi}{\begin{itemize}}
\newcommand{\ei}{\end{itemize}}
\newcommand{\ben}{\begin{enumerate}}
\newcommand{\een}{\end{enumerate}}

\def\ba{\begin{eqnarray}}
\def\ea{\end{eqnarray}}
\def\be{\begin{equation}}
\def\ee{\end{equation}}

\def\10{{\scriptscriptstyle 10}}
\def\4{{\scriptscriptstyle 4}}

\def\nn{\nonumber}
\def\mc{\mathcal}

\def\({\left(}
\def\){\right)}

\def\cD{{\cal D}}
\def\cE{{\cal E}}
\def\cF{{\cal F}}

\def\cO{{\cal O}}

\def\cV{{\cal V}}

\def\IZ{{\Bbb Z}}
\hypersetup{dvips,pdftitle={D3/D7 Branes at Singularities: Constraints from Global Embedding and Moduli Stabilisation}, pdfauthor={Michele Cicoli, Sven Krippendorf, Christoph Mayrhofer, Fernando Quevedo, Roberto Valandro}, pdfsubject={string phenomenology}, pdfkeywords={D-brane models, Calabi-Yau compactifications, moduli stabilisation, supersymmetry breaking}}

\title{D3/D7 Branes at Singularities: Constraints from Global Embedding and Moduli Stabilisation}

\author[a,b,c]{M.~Cicoli}
\author[d]{S.~Krippendorf}
\author[e]{C.~Mayrhofer}
\author[c,f]{F.~Quevedo}
\author[c,g]{R.~Valandro}

\affiliation[a]{Dipartimento di Fisica ed Astronomia, Universit{\`a} di Bologna, Bologna, Italy.}
\affiliation[b]{INFN, Sezione di Bologna, Italy.}
\affiliation[c]{Adbus Salam ICTP, Strada Costiera 11, Trieste 34014, Italy.}
\affiliation[d]{Bethe Center for Theoretical Physics and Physikalisches Institut der Universit\"at Bonn,\\
Nussallee 12, 53115 Bonn, Germany.}
\affiliation[e]{Institut f\"ur Theoretische Physik, Universit\"at Heidelberg,\\
Philosophenweg 19,  69120 Heidelberg, Germany.}
\affiliation[f]{DAMTP, University of Cambridge, Wilberforce Road, Cambridge, CB3 0WA, UK.}
\affiliation[g]{INFN, Sezione di Trieste, Italy.}

\abstract{In the framework of type IIB string compactifications on Calabi-Yau orientifolds we describe how to construct consistent global embeddings of
models with fractional D3-branes and connected `flavour' D7-branes at del Pezzo singularities
 with moduli stabilisation. 
Our results are applied to build an explicit compact example with a left-right symmetric model at a dP$_0$ singularity which features three families of chiral matter and gauge coupling unification at the intermediate scale. We show how to stabilise the moduli obtaining a controlled de Sitter minimum and spontaneous supersymmetry
breaking. We find an interesting non-trivial dynamical relation between the requirement of TeV-scale soft terms and the correct phenomenological values of the unified gauge coupling and unification scale.}

\preprint{DAMTP-2013-15}
\arxivnumber{1304.0022}

\begin{document}

\maketitle

\section{Introduction and summary}

D-branes at singularities in non-compact spaces have been much studied over the years
since they lead to promising Standard Model-like constructions with chiral matter, see~\cite{Malyshev:2007zz,Maharana:2012tu} for a review.
Given that most properties of these ultra-local models, in particular their matter content, are claimed to decouple from the gravitational physics of the bulk, their properties can be studied purely locally. There are infinite classes
of toric and non-toric singularities providing a rich spectrum of local models of particle physics
but it is not clear yet which of these models allow for a consistent global embedding in a proper string compactification.

We recently studied how to perform a consistent global embedding of local models in explicit compact Calabi-Yau (CY)
orientifolds \cite{Cicoli:2011qg,Cicoli:2012vw}. In \cite{Cicoli:2011qg} we addressed previously raised issues \cite{Blumenhagen:2007sm,Blumenhagen:2008zz,Collinucci:2008sq} describing how to combine moduli stabilisation with chirality
focusing on cases where the Standard Model-like sector is built with D7-branes wrapping divisors
in the geometric regime. On the other hand, in \cite{Cicoli:2012vw} we discussed the case of
fractional D3-branes at CY singularities without the inclusion of `flavour' D7-branes.
The main aim of this paper is to complete the case of D-branes at singularities, providing
a consistent global embedding of generic local models with both D3- and flavour D7-branes.\footnote{Flavour-D7 branes together with D3-branes at singularities were discussed in compact models but without moduli stabilisation in~\cite{Aldazabal:2000sa, Balasubramanian:2012wd}.}
\newpage
In \cite{Cicoli:2012vw} we succeeded in constructing type IIB string compactifications with
the following properties:
\begin{itemize}
\item Explicit description of the compact CY orientifold by means of toric geometry;

\item Chiral matter living on fractional D3-branes located at the singularities
obtained by collapsing two non-intersecting del Pezzo divisors mapped into each other by the
orientifold action;

\item A CY with at least one additional del Pezzo divisor
which is invariant under the orientifold involution such that it can support
a gauge theory that generates a non-perturbative superpotential for moduli
stabilisation;

\item On top of these three local four-cycles, there is at least one additional divisor
controlling the size of the CY volume, giving a total number of K\"ahler moduli $h^{1,1}\geq 4$;

\item Full classification of all models of this type from the Kreuzer-Skarke
list~\cite{Kreuzer:2000xy} of CY hypersurfaces in toric ambient spaces with $h^{1,2}\geq 5\geq h^{1,1}\geq 4$;

\item A visible sector gauge group including the Standard Model gauge symmetry;

\item Check of all consistency conditions including the cancellation of
D5- and D7-charges, Freed-Witten (FW) anomalies and K-theory torsion charges;\footnote{The non-vanishing D3-tadpole leaves enough space for background three-form fluxes to be turned on to stabilise the dilaton and complex structure moduli.}

\item Dynamical stabilisation of the K\"ahler moduli by considering both D- and F-term contributions
to the scalar potential in a way compatible with chirality;

\item Minkowski (or slightly de Sitter) vacuum for the closed string sector\footnote{Another type~IIB compact model with stabilised de~Sitter vacuum was found in \cite{Louis:2012nb} in the absence of chiral matter.} with supersymmetry
spontaneously broken by the F-terms of the K\"ahler moduli;

\item Generation of (sequestered) soft terms of order the TeV-scale for realistic matter on the D3-branes at del Pezzo singularities.
\end{itemize}
In this paper we extend this construction by providing consistent global embeddings of generic
local models with fractional D3-branes at singularities, `flavour' D7-branes
wrapping divisors which intersect the singularity, and bulk D7-branes which do not touch the singularity.
This gives rise to more generic models and allows us to obtain models with
spectra and couplings closer to the Standard Model than the cases with only
fractional D3-branes.

The flavour D7-branes have to wrap a holomorphic large cycle which intersects the blow-up divisor
resolving the singularity. Moreover, the restriction of the charges of these D7-branes
to the blow-up divisor has to yield the correct local charges of these flavour branes.
In the resolved picture the flavour D7-branes can look like branes wrapping connected or disconnected divisors. We study under which conditions the quiver system has only flavour branes that are connected in the resolved picture. This restriction allows us to deal with the flavour branes at large volume, study their effective field theory and hence analyse moduli stabilisation in this setup. We find that the restriction to this class of models that allow for an analysis of moduli stabilisation sets severe constraints on local model building.

We illustrate our general results by the particular case of a dP$_0$ singularity
embedded in an explicit CY three-fold. 
We choose a brane set-up such that the fractional branes at the singularity support
a left-right symmetric model with many interesting phenomenological features.\footnote{Notice that precisely this local left-right symmetric model was embedded globally in terms of an F-theory construction, albeit without moduli stabilisation, in \cite{Aldazabal:2000sa}. It would be interesting to study any connection of that embedding with the one presented in this paper.}
In fact, besides having three families of chiral matter, the gauge couplings
unify at the intermediate scale.

We show how to fix all the closed string moduli and
some of the open string scalars by a combination of D- and F-term contributions to the
scalar potential. Moduli stabilisation is performed systematically by classifying terms in the scalar potential in an expansion 
of inverse powers of the CY volume $\vo$~\cite{Balasubramanian:2012wd,Conlon:2005ki}.  The dilaton and complex structure moduli are as usual fixed by the leading $\mc{O}(\vo^{-2})$ terms generated by three-form fluxes and giving rise to the landscape of solutions which provide the two relevant parameters: the VEV of the tree-level flux superpotential $W_0$ and the string coupling $g_s$. Higher orders in the $\vo^{-1}$ expansion fix the K\"ahler moduli with exponentially large volume.
This potential has a structure which is rich enough to give rise to
Minkowski (or slightly de Sitter) vacua for a range of values of the underlying parameters $W_0$ and $g_s$.
These minima break supersymmetry spontaneously due to the presence of non-trivial
background fluxes which induce non-zero F-terms for some of the K\"ahler moduli.
In turn, soft terms are generated by gravity mediation.

Five important relations that characterise the phenomenological properties of our model are
the equations determining the minimum in the CY volume direction, the value of the cosmological constant $\Lambda$,
the energy scale of the soft terms $M_{\rm soft}$, the value of the string scale $M_s$ and of the
unified gauge coupling $\alpha^{-1}_{\rm unif}$. These five equations depend just on the two parameters
 $W_0$, $g_s$, and the value of the
CY volume $\vo$ at the minimum, giving rise to an over-determined system.
However, solving the first three equations, we find that the last two yield the desired phenomenological values,
providing a \textit{dynamical} explanation of gauge coupling unification.
In more detail, we fix $W_0\simeq 0.01$, $\vo \simeq 5\cdot 10^{11}$ and $g_s\simeq 1/65$ (within the perturbative regime)
by demanding the existence of a minimum of the scalar potential, $\Lambda = 0$ and $M_{\rm soft}\simeq 1$ TeV.
In turn, we obtain $M_s \simeq 10^{12}$ GeV which is the right energy scale where the gauge couplings unify,
and $\alpha^{-1}_{\rm unif} \simeq 20$ which is the exact value of the unified gauge coupling. Both 
of these physical quantities are determined independently from the low-energy spectrum and the RG running of the experimentally measured gauge couplings to higher energies in the left-right model at hand.

This paper is organised as follows. In section \ref{Sec2},
we first give a brief review of local brane models in non-compact CY backgrounds.
Then we describe how to embed them in a compact CY manifold with flavour D7-branes.
The consistency of the global embedding sets severe constraints on the local charges of flavour branes, as a consequence many local models cannot be realised with connected flavour branes.
Section~\ref{sec:example} illustrates these general results in an explicit example of a dP$_0$ quiver where we realise a left-right symmetric chiral model with three families of leptons and quarks, unification at an intermediate scale, Minkowski (or slightly de Sitter) moduli stabilisation and TeV-scale soft terms.
Finally we conclude in section~\ref{sec:conclusions}.

\section{Models with D3/D7 branes at a dP$_0$ singularity}
\label{Sec2}

\subsection{Non-compact models: a brief review}

The gauge theories associated to D-branes at non-compact singularities can be obtained for a vast set of geometric backgrounds, including toric singularities with and without flavour D7-branes~\cite{Douglas:2000qw,Franco:2006es,Yamazaki:2008bt} and non-toric singularities~\cite{Wijnholt:2002qz}. This class of supersymmetric gauge theories has proven to lead to phenomenologically viable and very interesting models, see for instance~\cite{Aldazabal:2000sa,Verlinde:2005jr,Conlon:2008wa,Krippendorf:2010hj,Dolan:2011qu} and~\cite{Maharana:2012tu} for a recent review. Within the local construction, in particular del Pezzo singularities have offered an interesting phenomenology which for example can account for hierarchical quark and lepton masses, Yukawa couplings leading to a realistic hierarchical flavour structure in the CKM and PMNS matrices. In addition, this class of models allows for gauge and matter extensions beyond the MSSM with potentially interesting phenomenology, but still consistent with gauge coupling unification, which could be observable at the LHC.

\begin{figure}
\begin{center}
 \includegraphics[width=0.4\textwidth]{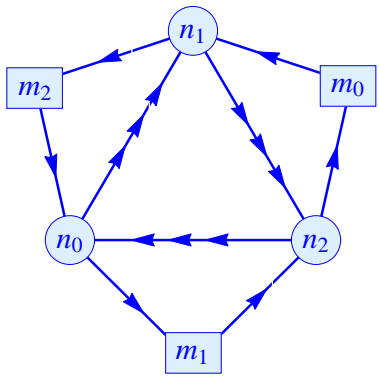}
 \end{center}
\caption{The dP$_0$ quiver encoding the $SU(n_0)\times SU(n_1)\times SU(n_2)$ gauge theory with flavour branes. Potential D7-D7 states are not shown.\label{fig:dp0quiverflav2}}
\end{figure}

Here we focus on the simplest models based on the zeroth del Pezzo singularity dP$_0$ or in other words the $\mathbb{C}^3/\IZ_3$ orbifold  singularity, as they already capture most of the characteristic features arising for D3/D7 branes at singularities. The extended quiver diagram including flavour D7-branes is shown in Figure~\ref{fig:dp0quiverflav2}.
Each node with label $n_i$ corresponds to a $U(n_i)$ gauge theory, arrows correspond to bi-fundamental fields $(n_i,\bar{n}_j).$ For each node, a distinct fractional brane exists. The $n_i$ denote the multiplicity of each fractional brane leading to the associated gauge group $U(n_i).$ Given a choice of D3 brane gauge groups $n_i,$ the flavour D7 brane gauge groups $m_0$, $m_1$, $m_2$  are constrained by anomaly cancellation:
\begin{equation}\label{m0m2asm1ni}
 m_0 = m + 3(n_1-n_0)\ , \qquad m_1=m\ ,\qquad m_2 = m + 3(n_1-n_2)  \:.
\end{equation}
Therefore, fixing the number of D3 branes $n_0$, $n_1$ and  $n_2$, in order to look for a realistic model, determines the number of D7 branes up to a free integer $m.$\footnote{The numbers $m_i$ do not necessarily imply $U(m_i)$ gauge symmetries but can be, for instance, products of $U(1)$ gauge symmetries and instead of one single arrow connecting the D3 and D7 branes there may be multiple arrows with reduced gauge symmetry, all this is encoded in the choice of $m_i$ determined by anomaly cancellation.}

For example, choosing all D3-brane gauge groups to equal three, $n_0=n_1=n_2=3$, leads to the trinification model, and the choice $n_0=n_2=2$,  $n_1=3$ leads to a left-right extension of the Standard Model gauge group. Regarding flavour branes, the trinification model allows for additional D7 branes with universal $m_i=m$, whereas  in the left-right model D7-branes are required for anomaly cancellation. The quiver diagram of the minimal choice for D7-branes in the left-right model is shown in Figure~\ref{fig:dp0quiver}.

Note that this arbitrariness with respect to the choice of flavour branes provides more flexibility for model building, i.e.~it opens a previously un-accessible but very interesting class for model building. In addition, we note that, as for instance in the trinification model, generically the flavour D7 branes provide alternative options to break the symmetry group to the Standard Model one.
As discussed in~\cite{Cicoli:2012vw}, the trinification model with no flavour branes ($m = 0$) requires an involved two-step breaking procedure where the right-handed sneutrino does the breaking to the Standard Model gauge group. One could avoid this complication by introducing flavour D7-branes in the local model. 

Notice also that models with different values of $n_0$, $n_1$, $n_2$ are in principle disconnected gauge theories. In other words, one single singularity, like dP$_0,$ gives rise to an infinite number of possibly unconnected local chiral models by changing values of $n_i$ and $m$. We show in~\cite{Cicoli:2013zha} that a consistent global embedding of these models allows for transitions from one set of values for $n_i$, $m$ to a different one.

\subsection{Compact models with flavour branes: constraints from global embedding} \label{modifsect}

We now want to embed the local models on the dP$_0$ singularity with D3 and flavour D7-branes described above in a compact CY manifold. 

If there are distinct numbers of fractional D3-branes\footnote{A D3-brane at a $\mathbb{C}^3/\mathbb{Z}^3$ singularity can be roughly seen as a collection of fluxed D7-branes wrapping the shrinking divisor. More precisely, a D3-brane at a singularity splits into a collection of fractional branes. The fractional branes are described by coherent sheaves ${\cal E}$ on the (shrinking) dP$_0$ divisor.} $n_i\neq n_j$, we have $m_i\neq 0$ for some $i$, cf.~\eqref{m0m2asm1ni}, and hence some flavour brane exists. This is a D7-brane that wraps a large four-cycle in the compact geometry and passes through the dP$_0$ singularity, i.e.~it intersects the dP$_0$ in the blown-up picture.

To have a consistent global embedding, one needs to check that the local D-brane charges of the flavour D7-branes
come from the restriction of the charges of globally well-defined D7-branes wrapping a holomorphic divisor
of the compact CY $X$ to the dP$_0$.

The RR charges of both D7-branes and fractional D3-branes are formally expressed by the `Mukai' charge vectors of
D-branes.
The D-brane charges of a configuration of D-branes are given by the sum of their Mukai vectors which are defined by
\begin{equation}
 \Gamma_{{\cal E}} = D\wedge \mbox{ch} ({\cal E}) \wedge \sqrt{\frac{\mbox{Td}(TD)}{\mbox{Td}(ND)}} \:,  \qquad
   \mbox{ with } \qquad S_\textmd{D-br} = \int_{\mathbb{R}^{1,3}\times X} C \wedge e^{-B} \wedge \Gamma_{{\cal E}}.
\label{ChargeVectD7br}
\end{equation}
Here $D$ is the Poincar\'e dual of cycle wrapped by the D-brane\footnote{
In this article, we will use the same symbol for the cycles and their Poincar\'e dual forms.
}, $B$ is the NS B-field, $C$ is the formal sum of the RR p-form potentials,
Td$(V)= 1 + \frac12 c_1(V) + \frac{1}{12}(c_1(V)^2+c_2(V))+... $ is the Todd class of the vector bundle $V$,
$TD$ is the tangent bundle of $D$ and $ND$ the normal bundle of $D$ in $X$ while
ch$({\cal E})$ is the Chern character of the vector bundle ${\cal E}$, more precisely a sheaf, living on the brane.\footnote{The charge vector can also be written in terms of the A-roof genus $\hat{A}$, by shifting the sheaf $\cE$ to the sheaf ${\cal W}={\cal E}\otimes K_S^{1/2}$ whose first Chern class is identified with the gauge flux.}
Looking at $S_\textmd{D-br}$ in \eqref{ChargeVectD7br}, one finds that the D7-charge is encoded in the two-form component of $e^{-B}\Gamma_{\cal E}$,
the D5-charge in the four-form and the D3-charge in the six-form.\footnote{These p-forms are actually the push-forward to the CY manifold $X$ of forms on the D-brane, for a review see \cite{Aspinwall:2004jr}. For this reason, a two-form flux on a D7-brane, Poincar\'e dual to a curve $C$ that is trivial on $X$ but non-trivial on the D7-brane, will appear in the D3-charge (six-form) but not in the D5-charge (four-form).}

Using~\eqref{ChargeVectD7br} and the fact that $X$ is a CY, we obtain for a D7-brane which wraps the divisor class $D$ and has abelian gauge flux $\cF$
\be\label{ChVectExp}
 \Gamma_{D7}(D,{\cal F})\equiv  e^{-B}\Gamma_{{\cal E}} = D \left( 1 + {\cal F} + \frac12 {\cal F}\wedge {\cal F} + \frac{c_2(D)}{24} \right),
\ee
where ${\cal F}=F-B$ and $F=c_1({\cal E})+\frac{c_1(D)}{2}$. From this we can read off the RR charges of the D7-brane.
The charge vector of the image D7-brane D7$'$, wrapping the image divisor $D'$ is given by $\Gamma_{D7}(D',-{\cal F'})$.

We start by considering fractional branes at the dP$_0$ singularity.
A fractional brane corresponds to a bound state described by a coherent sheaf $F_a$ on the dP$_0$ surface; it is characterised by the charge vector of a D-brane wrapping the shrinking cycle.
For a dP$_0$ singularity, one has three types of mutually stable branes, referred to as fractional branes.
The geometric part of the vector --- the square root part of $\Gamma$ in \eqref{ChargeVectD7br} ---
is the same for all of them,
as they wrap the same divisor. The vector bundle (sheaf) part is different for the three fractional branes and is given by~\cite{Diaconescu:1999dt,Douglas:2000qw}\footnote{Note that we use the opposite sign convention with respect to the literature on D-branes at dP$_n$ singularities. This is because in our convention a D7(anti-D7)-brane has charge $+1(-1)D$, where $D$ is the holomorphic wrapped divisor. Note that in this convention, the D3-charge is minus the integral of the six-form component of $\Gamma_{D7}$.}
\be
 \mbox{ch}({F_0}) = -1 + H -\tfrac12\, H\wedge H\,,  \quad
 \mbox{ch}({F_1}) = 2 - \,H -\tfrac12\, H\wedge H\,, \quad
 \mbox{ch}({F_2}) = - 1\,.
\label{fluxsheaf}
\ee
Since $b_2({\rm dP}_0)=1$, all the divisors on a dP$_0$ are proportional to the hyperplane class $H$
that generates $H^{1,1}({\rm dP}_0)$. Its Poincar\'e dual two-form will be the pullback of a two-form of the CY $X$; we call $D_H$ this two-form and its Poincar\'e dual divisor in $X$. Note there is an ambiguity in choosing $D_H$, as we can add to it any two-form of $X$ whose pullback onto the dP$_0$ is trivial.

We can now compute the global charge vectors \eqref{ChargeVectD7br} for the three fractional branes wrapping the shrinking divisor $\cD_{\textmd{dP}_0}$:
\begin{eqnarray}\label{FractionalD3ChVect}
 \Gamma_{F_0} &=& \cD_{\textmd{dP}_0}\wedge\left\{ -1  -\tfrac12 D_H  -\tfrac14  D_H \wedge D_H  \right\}\,, \nonumber\\
 \Gamma_{F_1} &=& \cD_{\textmd{dP}_0}\wedge\left\{ 2 +2  D_H + \tfrac12  D_H \wedge D_H  \right\}\,, \\
 \Gamma_{F_2} &=& \cD_{\textmd{dP}_0}\wedge\left\{ -1 - \tfrac32 D_H -\tfrac54  D_H \wedge D_H  \right\}\,.\nonumber
\end{eqnarray}

From the charge vector, one can also compute the number of chiral states in the bi-fundamental representation between the different nodes of the quiver. It is given by the following product of the two Mukai vectors:\footnote{Given two branes $D7_1$ and $D7_2$, $n=\langle \Gamma_{D7_1},\Gamma_{D7_2} \rangle>0$ means that we have $n$ states which are in the anti-fundamental representation of the $D7_1$ gauge group and in the fundamental representation of the $D7_2$.}
\begin{equation}
 \langle \Gamma_1,\Gamma_2 \rangle \equiv \int_X \left( - \Gamma_1^{\rm (2-form)} \wedge \Gamma_2^{\rm (4-form)}
    + \Gamma_2^{\rm (2-form)} \wedge \Gamma_1^{\rm (4-form)}  \right)
\end{equation}
where $\Gamma^{\rm (n-form)}$ is the n-form component of the charge vector $\Gamma$. Applying this formula to the charge vectors in \eqref{FractionalD3ChVect} we obtain the known result about the dP$_0$ quiver, i.e.\ that the number of chiral states between each pair of nodes is equal to three:
\begin{equation}
   \langle \Gamma_{F_1} , \Gamma_{F_0}\rangle =    \langle \Gamma_{F_2} , \Gamma_{F_1}\rangle =    \langle \Gamma_{F_0} , \Gamma_{F_2}\rangle = 3 \:.
\end{equation}

Let us move to the flavour D7-branes. Each of them will have an associated charge vector
\begin{equation}
 \Gamma_{\cD_{\rm flav}} = \cD_{\rm flav}\wedge \mbox{ch} (\cE_{\rm flav}) \wedge \sqrt{\frac{\mbox{Td}(T\cD_{\rm flav})}{\mbox{Td}(N\cD_{\rm flav})}} \:,
\label{ChVectFlav}
\end{equation}
where $\cD_{\rm flav}$ is the divisor wrapped by the globally defined flavour brane and $\cE_{\rm flav}$ is the vector bundle living on it. Differently to the fractional branes, the global flavour brane charge vectors are not fully determined by the local model we want to embed. The reason is that the flavour D7-brane extends in the non-compact directions. 
The only information that the local model gives is the number of chiral intersections between the flavour and the fractional branes which depend on the flavour brane D7- and D5-charges restricted to the dP$_0$. These are encoded in the local charge vector of the flavour brane defined by the pullback of the global charge vector to dP$_0$:
\begin{equation}\label{LocalD7flCharg}
  \Gamma_{D7_i}^{\rm loc} \equiv \left.\Gamma_{\cD_{\rm flav}^i} \right|_{\textmd{dP}_0} = a_i H + b_i H \wedge H  \qquad \qquad \mbox{with } i=0,1,2\:,
\end{equation}
where $i$ runs over the three type of flavour branes associated to the nodes $m_i$ of the quiver diagram in Figure~\ref{fig:dp0quiverflav2}. In \eqref{LocalD7flCharg} we have used the fact that any divisor class restricted to the dP$_0$ is a multiple of the hyperplane class $H$.

The coefficients $a_i$ and $b_i$ in \eqref{LocalD7flCharg} are determined by requiring the right amount of chiral states to make the full quiver system anomaly free. As we see in Figure~\ref{fig:dp0quiverflav2}, in the used conventions the $i$-th flavour brane does not have any chiral intersection with the $i$-th fractional brane, leading to the following constraints on $a_i$ and $b_i:$
\begin{eqnarray}
\left. \begin{array}{ccl}
m_0 &=& \langle \Gamma_{D7_0}^{\rm loc},\Gamma_{F_2} \rangle = \frac32 a_0 - b_0 \\
 0 &=& \langle \Gamma_{D7_0}^{\rm loc},\Gamma_{F_0} \rangle = \frac12 a_0 - b_0\\
\end{array} \right\}  &\,\,\Rightarrow\,\,& a_0=m_0 \qquad b_0=\frac{m_0}{2}\,, \\
\left. \begin{array}{ccl}
m_1 &=& \langle \Gamma_{D7_1}^{\rm loc},\Gamma_{F_0} \rangle = \frac12 a_1 -b_1 \\
 0 &=& \langle \Gamma_{D7_1}^{\rm loc},\Gamma_{F_1} \rangle = -2 a_1 + 2 b_1\\
\end{array} \right\}  &\,\,\Rightarrow\,\,& a_1= -2 m_1 \qquad b_1= -2 m_1\,, \\
\left. \begin{array}{ccl}
m_2 &=& \langle \Gamma_{D7_2}^{\rm loc},\Gamma_{F_1} \rangle = -2a_2 + 2b_2 \\
 0 &=& \langle \Gamma_{D7_2}^{\rm loc},\Gamma_{F_2} \rangle = \frac32 a_2 - b_2\\
\end{array} \right\}  &\,\,\Rightarrow\,\,& a_2=m_2 \qquad b_2=\frac{3m_2}{2}\,.
\end{eqnarray}
By using \eqref{m0m2asm1ni}, the local flavour brane charge vectors are
\begin{eqnarray}\label{GenericFlavChVct}
 \Gamma_{D7_0}^{\rm loc} &=& (m+ 3(n_1-n_0)) H \left( 1+ \frac12 H\right) \,,\nonumber\\
 \Gamma_{D7_1}^{\rm loc} &=& -2m H \left( 1+ H\right) \,,\\
 \Gamma_{D7_2}^{\rm loc} &=& (m+ 3(n_1-n_2)) H \left( 1+ \frac32 H\right)\,.\nonumber
\end{eqnarray}

We now require these local vectors to come from the restriction of the Mukai vectors \eqref{ChVectFlav} of consistent D7-branes. 
The flavour D7-brane wraps a divisor $\cD_{\rm flav}$ that passes through the singularity where the fractional D3-brane sits. To preserve supersymmetry this divisor must be holomorphic.
In the resolved picture, i.e.\ when the singularity is blown up and replaced by an exceptional divisor $\cD_{\textmd{bu}}$ which in the case of $\mathbb C^3/\mathbb Z_3$ is a ${\textmd{dP}_0}$,
the homology class of the divisor wrapped by the flavour brane has to satisfy the condition
\[
\cD_{\rm flav}|_{\cD_{\textmd{dP}_0}}=a_i H,
\]
where $a_i$ is given by the local model we want to embed. 
We now see that depending on the sign of $a_i$, the flavour brane is either a connected or disconnected object in the resolved picture. In the latter case, some quantities, such as the FI-terms, 
are more involved 
to determine by using large volume techniques.
For this reason, we will stick in our analysis to the first case where all the flavour branes are irreducible divisors in the resolved picture.

When the homology class of $\cD_{\rm flav}$ has a connected (i.e.\ non-factorised) element, 
the intersection of the 
divisor $\cD_{\rm flav}$ with the blow-up divisor $\cD_{\textmd{bu}}$ must be some effective curve ($\cD_{\rm flav} \cap \cD_{\textmd{bu}}\ne \emptyset$). 
This curve, therefore, lies in the Mori cone of the resolution divisor $\cD_{\textmd{bu}}$ --- correspondingly the Poincar\'e dual two-form lies in its K\"ahler cone. Since we want chiral modes between the flavour D7-brane and the fractional branes, the intersection curve must be in a homology class whose push-forward is in a non-trivial homology class of $X$. As we have seen, the D7-brane charge is given by the two-form Poincar\'e dual to the divisor wrapped by the D7-brane, cf.~\eqref{ChVectExp}. The locally induced D7-charge is the pullback of this two-form to the blown-up divisor $\cD_\textmd{bu}$. However,
this is just the two-form Poincar\'e dual to the (effective) intersection curve $\cD_{\rm flav} \cap \cD_{\textmd{bu}}$. Hence, 
\textit{the two-form encoding the locally induced D7-brane charge of a `connected' flavour brane must lie in the K\"ahler cone of the blow-up divisor.}
%
For the case at hand, the class of the intersection curve is a {\it positive} multiple of the hyperplane class of $\mathbb P^2$.\footnote{The effective curve is the intersection of two holomorphic divisors in a three-fold. An effective curve on a dP$_0$ is given by the vanishing locus of a homogeneous polynomial of degree $n>0$. The class of this curve is then $n\,H$ and therefore it is a positive multiple of $H$. The Mori cones for dP$_n$ surfaces with $n>0$ can, for instance, be found in Appendix A of~\cite{Beasley:2008dc}.} Therefore the local D7-brane charge has to be a positive multiple of~$H$. If it were not positive, then the divisor $\cD_{\rm flav}$ would be forced to split into two components, one of which would be (a multiple of) $\cD_{\textmd{dP}_0}$ ($\cD_{\textmd{bu}}$) itself. 
Therefore, \emph{a flavour brane with a negative local D7-charge looks, in the resolved picture, like a disconnected object: one piece is a connected large volume four-cycle, while the other component is a multiple of $\cD_{\rm {bu}}$.} As suggested in~\cite{Balasubramanian:2012wd}, when the divisor $\cD_{\textmd{bu}}$ is shrunk to zero size, $\alpha'$ effects should be responsible for the formation of one bound state, i.e.~the flavour brane, out of the two components.
In Appendix~\ref{appendix}, we show that this fact is already present in the non-compact $\mathbb{C}^3/\mathbb{Z}_3$ case.

This reasoning gives strong constraints on the numbers $n_i$ and $m_i$ for the local model with only `connected' flavour branes:
\begin{equation}\label{nogonm}
 -2m\geq 0\ , \qquad 3(n_1-n_0)+m \geq 0\ , \qquad 3(n_1-n_2)+m \geq 0\:,
\end{equation}
which is equivalent to
\begin{equation}\label{nogonmbis}
  0 \leq -m \leq 3(n_1 - \max\{n_0,n_2\})\,.
\end{equation}
In particular, we see that these conditions imply $n_1\geq n_0$ and $n_1\geq n_2$. We also note that when $n_1=n_2=n_3$ we obtain $m=0$ and consequently also $m_0=m_1=m_2=0$. 


These constraints are not manifestly invariant under the $\mathbb{Z}_3$ symmetry of the quiver. If we apply a $\mathbb{Z}_3$ rotation to a quiver system satisfying the constraints \eqref{nogonmbis},  we can end up with a system that violates them. 
For our phenomenological constructions we will only use quiver theories 
satisfying \eqref{nogonmbis} up to a $\mathbb{Z}_3$ rotation and we will fix the $\mathbb{Z}_3$ symmetry such that all the flavour branes are connected in the resolved picture.
However, models that do not satisfy \eqref{nogonmbis} up to a $\mathbb{Z}_3$ rotation, will always have disconnected flavour branes in the resolved picture. 

Again, for model building we restrict ourselves to setups with `connected' flavour branes as they allow for a straightforward description in terms of effective field theory which is necessary for addressing moduli stabilisation.

\section{Explicit example of a dP$_0$ quiver}
\label{sec:example}

We now present an explicit example of a globally embedded quiver gauge theory with flavour branes, that satisfies the constraints \eqref{nogonmbis}. We consider the same CY three-fold $X$ that we have used in~\cite{Cicoli:2012vw}. This was chosen from a list of hypersurfaces in toric ambient varieties that satisfy certain requirements:
the CY must have $h^{1,1}\geq 4$, two independent dP$_n$ divisors and one further rigid divisor; moreover, there must be an (orientifold) involution which exchanges the two dP$_n$ divisors  such that they do not include fixed points. When the two dP$_n$ are shrunk to zero size we obtain two singularities exchanged by the orientifold involution and an orientifold plane that does not pass through the singular point. If we put a suitable set of fractional branes on top of the two singularities we can realise the desired quiver gauge theory in the compact CY manifold. Note that in the physical space (after the orientifold quotient) we have only one quiver model.

\subsection{Geometric setup}

In this section we summarise the details of the chosen  CY manifold $X$, see~\cite{Cicoli:2012vw} for more details.
$X$ is a hypersurface in the toric ambient variety defined by the following weight matrix and Stanley-Reisner ideal
\begin{equation}
\begin{array}{|c|c|c|c|c|c|c|c||c|}
\hline z_1 & z_2 & z_3 & z_4 & z_5 & z_6 & z_7 & z_8 & D_{eq_X} \tabularnewline \hline \hline
    1  &  1  &  1  &  0  &  3  &  3  &  0  &  0  & 9\tabularnewline\hline
    0  &  0  &  0  &  1  &  0  &  1  &  0  &  0  & 2\tabularnewline\hline
    0  &  0  &  0  &  0  &  1  &  1  &  0  &  1  & 3\tabularnewline\hline
    0  &  0  &  0  &  0  &  1  &  0  &  1  &  0  & 2\tabularnewline\hline
\end{array}
\label{eq:model3dP0:weightm}\,
\end{equation}
\begin{equation}
{\rm SR}=\{z_4\, z_6,\,z_4\, z_7, \, z_5\, z_7,\, z_5\, z_8,\, z_6\, z_8,\, z_1\, z_2\, z_3\}\,.\nonumber
\end{equation}
In~\eqref{eq:model3dP0:weightm} the last column refers to the degrees of the hypersurface equation $eq_X=0$. The Hodge numbers of the CY are $h^{1,1}=4$ and $h^{1,2}=112$, such that $\chi=-216$.
Furthermore, each of the three toric divisors $D_4$, $D_7$ and $D_8$ corresponds to a $\mathbb P^2=$dP$_0$, on $X$. They do not mutually intersect.

For $H^{1,1}(X)$ we choose the basis\footnote{Note that this basis of integral cycles is not an `integral basis'; in particular $D_1=\frac{1}{3}(\cD_b-\cD_{q_1}-\cD_{q_2}-\cD_s)$.}
\be
\cD_b = D_4+D_5=D_6+D_7,\qquad \cD_{q_1} = D_4,\qquad \cD_{q_2} = D_7,\qquad \cD_s = D_8\,,
\label{eq:simplebasis}
\ee
where `$b$' refers to `big' since it controls the overall size of the CY,
`$q_i$' $i=1,2$ for `quiver' since these will shrink to dP$_0$-singularities
exchanged by the orientifold action, and `$s$' for `small' since this divisor will support
non-perturbative effects with size much smaller than the large four-cycle.
The intersections between the basis elements take a simple form
\be
\label{intersnumb}
I_3 = 27\,\cD_b^3 + 9\,\cD_{q_1}^3 + 9\,\cD_{q_2}^3 + 9\,\cD_s^3 \:.
\ee
Expanding the K\"ahler form in the basis \eqref{eq:simplebasis} as $J= t_b \cD_b + t_{q_1}\cD_{q_1} + t_{q_2}\cD_{q_2}+t_s\cD_s$,
the volumes of the four divisors become ($\tau_i := {\rm Vol}(\cD_i)=\frac 12 \int_{\cD_i} J\wedge J$)
\be
\tau_b = \frac{27}{2}\, t_b^2\,,
\qquad \tau_{q_1}  = \frac 92 \, t_{q_1}^2 \,,
\qquad \tau_{q_2} = \frac 92 \, t_{q_2}^2 \,,
\qquad \tau_s = \frac 92 \, t_s^2\,.
\ee
The diagonal structure is also reflected in the `Swiss-cheese' form of the CY volume
\be
\vo := {\rm Vol}(X) = \frac 32 \left(3 t_b^3 + t_{q_1}^3 + t_{q_2}^3 + t_s^3\right)
=\frac{1}{9}\sqrt{\frac{2}{3}}\left[\tau_b^{3/2} -  \sqrt{3}\left( \tau_{q_1}^{3/2} + \tau_{q_2}^{3/2} + \tau_s^{3/2}\right)\right]\,,
\label{vol}
\ee
where ${\rm Vol}(X) = \frac 16 \int_X J\wedge J\wedge J.$

\subsubsection*{Orientifold involution}

The orientifold involution that exchanges two of the three dP$_0$ divisors is
\be
\label{eq:OrInvol}
 z_4\leftrightarrow z_7\qquad\textmd{and}\qquad z_5\leftrightarrow z_6\:.
\ee
The CY hypersurface $X$ must be symmetric under this holomorphic involution: its complex structure must be then such that the defining equation $eq_X=0$ is symmetric under the involution.
From~\eqref{eq:OrInvol} we see that the two del Pezzo surfaces at $z_4=0$ ($\cD_{q_1}$) and $z_7=0$ ($\cD_{q_2}$) are interchanged by this involution. Furthermore, in \cite{Cicoli:2012vw} we showed that the fixed locus of \eqref{eq:OrInvol} is given by the following two orientifold planes:
\begin{equation*}
\begin{array}{ccccc}
 \mbox{O7-planes}\quad&&\quad \mbox{Locus in ambient space}\quad &&\quad \mbox{Homology class } \\
   O7_1: && y_6=z_4z_5-z_6z_7=0 && D_6+D_7=\cD_b\,, \\ O7_2: &&   y_5=z_8=0 && D_8=\cD_s\,. \\
\end{array}
\end{equation*}

\subsubsection*{K\"ahler cone and relevant volumes}

In the smooth case, the integral of $J$ over all effective curves of $X$ has to be positive definite, i.e.~$\int_{\mc{C}_j}J>0$
for all curves $\mc{C}_j$ in the Mori cone of $X$. This defines the K\"ahler cone of $X$. After a subtle analysis performed in \cite{Cicoli:2012vw}, one finds the following K\"ahler cone conditions on the coefficients $t_i$
\be
t_b+t_{q_1}>0\,, \qquad t_b+t_{q_2}>0\,, \qquad t_b+t_s>0\,, \qquad t_{q_1}<0\,, \qquad t_{q_2}<0\,, \qquad t_s<0\,. \nn
\ee
Under the orientifold involution, the K\"ahler form is even and must therefore belong to $H^{1,1}_+(X)$.
This is obtained by taking $t_{q_1}=t_{q_2}$.
Moreover, we want the two dP$_0$ divisors at $z_4=0$ and $z_7=0$ to shrink to zero size
in order to generate the two (exchanged) dP$_0$ singularities.
This is realised on the boundary of the K\"ahler cone given by $t_{q_1}=t_{q_2}=0$.
The remaining K\"ahler cone conditions are
$$t_b+t_s >0 \qquad \textmd{and}\qquad t_s < 0\,.$$

\subsection{Global embedding with flavour branes}
\label{sec:GlobalEmbdP0}

We consider the case when the two dP$_0$ divisors, $\cD_{q_1}$ and $\cD_{q_2}$, are collapsed to zero size, generating two $\mbb{C}^3/\mbb{Z}_3$ singularities, while the other dP$_0$ divisor $\cD_s$ is of finite size.
As we have seen in \cite{Cicoli:2012vw}, the vanishing of the two K\"ahler moduli $\tau_{q_1}=\tau_{q_2}$
is enforced by D-terms and we discuss this detail further in section~\ref{sec:modstab}.
\begin{figure}
\begin{center}
\includegraphics[width=0.45\textwidth]{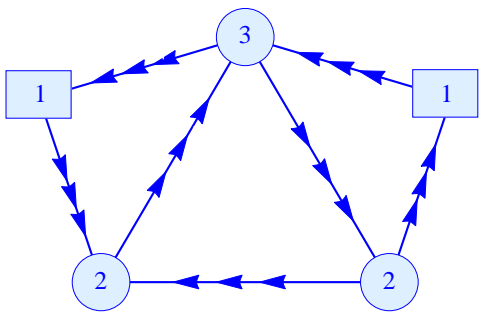}
\end{center}
\caption{The dP$_0$ quiver encoding the $SU(3)\times SU(2)^2$ gauge theory with flavour branes. Again only D3-D3 and D3-D7 states are shown.}\label{fig:dp0quiver}
\end{figure}

\subsubsection{Brane set-up, fluxes and chiral spectrum}
\label{BraneSetUp}

We have the following set of O-planes and D-branes:
\begin{itemize}
\item There are two orientifold O7-planes, one at $z_4z_5-z_6z_7=0$,
lying in the class $\cD_b$,
and the other at $z_8=0$, in the class $\cD_s$.
The two fixed loci are disconnected and do not intersect each other.

\item We put the system of fractional branes, shown in Figure~\ref{fig:dp0quiver}, on the singularity at $z_4=0$ and their images on the singularity at $z_7=0$  to have an invariant configuration, cf.\ Figure~\ref{fig:setup}.
\begin{figure}
\begin{center}
\includegraphics[width=0.5\textwidth]{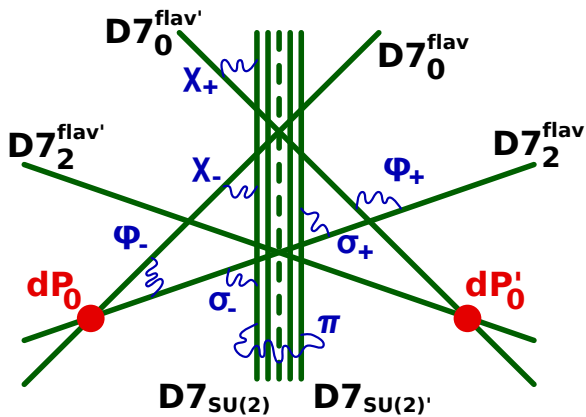}
\end{center}
\caption{Brane setup: The red points represent the fractional branes. There are two branes (plus their images) on top of the O-plane (dotted line) and two flavour branes (plus their images). The fields $\varphi_{\pm}$, $\chi_{\pm}$, $\sigma_{\pm}$, $\pi$ are the chiral modes that are generated in this construction by the chosen fluxes. The non-perturbative cycle is neglected (it does not intersect the other cycles).}
\label{fig:setup}
\end{figure}
The visible sector is given by the fractional D3-branes with gauge theory $SU(3)_c\times {SU(2)}_L\times SU(2)_R\times U(1)_{ \textmd{B-L}}$.\footnote{There are two anomalous $U(1)$ symmetries which become massive by eating up
the local axions given by the reduction of the RR forms $C_4$ and $C_2$
on the dP$_0$ divisor and its dual two-cycle (the hyperplane class of $H^{1,1}({\rm dP}_0)$).
The remaining $U(1)_{\textmd{B-L}}$ factor is an anomaly-free and massless $U(1)$ symmetry.}

\item We will need some D7-branes playing the r\^ole of the flavour D7-branes in the quiver diagram, and further stacks of D7-branes that saturate the D7-tadpole but do not intersect the shrinking dP$_0$ surfaces, cf.\ Figure~\ref{fig:setup}.

\end{itemize}

The visible sector is given by the gauge group on the fractional branes and its chiral matter.
The D7-branes wrapping the fixed locus at $z_8=0$ gives a pure $SO(8)$ Yang-Mills theory which undergoes gaugino condensation. In fact, no zero modes are generated due to the fact that the four-cycle $\cD_s$ is rigid and does not intersect the cycles wrapped by the other D7-branes in the configuration.

\subsubsection*{D-brane charges of the quiver gauge theory}

We start by considering the fractional branes at the two dP$_0$ singularities at $z_4=0$ and $z_7=0$.
From the intersection number $D_4 \cap D_1 \cap D_1 = 1$, we see that $D_1|_{D_4} = H$. Hence, the $D_H$ divisor is given by $D_1$ modulo a linear combination of $\cD_b$, $\cD_s$ and $\cD_{q_2}$.
The global charge vectors \eqref{ChargeVectD7br} for the three fractional branes wrapping the locus at $z_4=0$ are then
\begin{eqnarray}\label{FractionalD3ChVectExample}
 \Gamma_{F_0} &=& \cD_{q_1}\wedge\left\{ -1  -\tfrac12 D_1  -\tfrac14  D_1 \wedge D_1  \right\}\,, \nonumber\\
 \Gamma_{F_1} &=& \cD_{q_1}\wedge\left\{ 2 +2  D_1 + \tfrac12  D_1 \wedge D_1  \right\}\,, \\
 \Gamma_{F_2} &=& \cD_{q_1}\wedge\left\{ -1 - \tfrac32 D_1 -\tfrac54  D_1 \wedge D_1  \right\}\,.\nonumber
\end{eqnarray}
The quiver diagram in Figure~\ref{fig:dp0quiver} corresponds to taking the following multiplicities
\begin{equation}\label{multExamp}
  n_0 =2\ , \qquad\qquad n_1 =3\ , \qquad\qquad n_2 =2\,.
\end{equation}
We see that they satisfy condition \eqref{nogonm} with $-3\leq m\leq 0$. In the following we will choose $m=0$. The case $m=-3$ corresponds to having only flavour branes of type $D7_1^{\rm flav}$ and can be obtained by recombining the two flavour branes $D7_0^{\rm flav}$ and $D7_2^{\rm flav}$ in Figure~\ref{fig:setup}.

The total charge vector of the fractional branes at $z_4=0$ is
\begin{eqnarray}
 \Gamma_{{\rm frac D3}^{(1)}} &=& 2 \Gamma_{F_0} + 3 \Gamma_{F_1} + 2 \Gamma_{F_2}  =
    \cD_{q_1}\wedge\left\{ 2 +2 D_1 - \tfrac32  D_1 \wedge D_1  \right\} \, .
\end{eqnarray}
The same happens for the fractional branes at $z_7=0$. Since we want an orientifold invariant configuration, we need to take the same multiplicities for the fractional branes.
Notice that we have a net non-zero D7-brane charge: roughly, it is the charge of two D7-branes wrapping the shrinking divisor.

From \eqref{GenericFlavChVct} and \eqref{multExamp}, one can compute the local charges of the flavour D7-branes in Figure~\ref{fig:setup}. We obtain
\begin{equation}\label{localD7flch}
  \Gamma_{D7_0}^{\rm loc} = 3H ( 1 + \frac12 H)\,, \qquad\qquad  \Gamma_{D7_2}^{\rm loc} = 3H ( 1 + \frac32 H) \:.
\end{equation}
Note that these charges realise local D7- and D5-charge cancellation, as expected from anomaly cancellation.
To check this we need to know that the pullback of $D_4$ to the shrinking dP$_0$ is given by $-3H$ because $D_4$ is a $\mathbb P^2$.

The relations \eqref{localD7flch} mean that both flavour branes must wrap a divisor class in $X$ whose Poincar\'e dual two-form gives $3H$ once pulled back to dP$_0$.
We have some ambiguity in choosing this class. In fact, any combination $3D_1 + \alpha^b\cD_b + \alpha^s \cD_s + \alpha^{q_2}\cD_{q_2}$ restricts to $3H$ on $\cD_{q_1}$, since $\cD_b$, $\cD_s$ and $\cD_{q_2}$ are trivial when pulled-back to $\cD_{q_1}$.
We fix $\alpha^s$ to one by requiring zero intersection with the cycle $\cD_s$ supporting non-perturbative effects. We fix $\alpha^{q_2}$ to one by demanding that the flavour branes of the quiver system at $z_4=0$ do not intersect the image (shrinking) dP$_0$ $\cD_{q_2}$ (in fact $(3D_1+\cD_{q_2})\cdot \cD_{q_1}=0$).\footnote{
This second condition is not necessary for phenomenology. We might also choose $\alpha_{q_2}=0$: in this case, the brane wrapping $3D_1+\cD_s+\alpha^b\cD_b$ would be a flavour brane for both the quiver system at $z_4=0$ and its image at $z_7=0$. In particular, being an invariant brane, its flux must be odd under the orientifold involution and the FI-term would then vanish.}
The classes of the flavour branes are then
\begin{eqnarray*}
  \label{D7flChVect1}{[D7^{\rm flav}_0]}&=&3D_1 + \cD_s +\cD_{q_2} + \alpha_0^b\cD_b=(1+\alpha_0^b)\cD_b-\cD_{q_1} \,,\\
  \label{D7flChVect2}{[D7^{\rm flav}_2]}&=&3D_1 + \cD_s +\cD_{q_2} + \alpha_2^b\cD_b=(1+\alpha_2^b)\cD_b-\cD_{q_1}\,.
\end{eqnarray*}
We will choose in the following $\alpha_0^b=\alpha_2^b=0$. In this case,
the two branes wrap two different representatives of the same homology class $\cD_b-\cD_{q_1}$. These divisors are connected surfaces inside the CY $X$, whose equations are generically of the form $z_5+P_i^3(z_1,z_2,z_3)\,z_7\,z_8$, where $P_i^3(z_1,z_2,z_3)$ ($i=1,2$) are two polynomials of degree three in the coordinates $z_1$, $z_2$, $z_3$.

The gauge flux living on the flavour branes is encoded in the four-form of the charge vector.
As can be seen from the expansion of \eqref{ChVectFlav}, this four-form is $\cD_{\rm flav}\wedge \cF_{\rm flav}$ (we will consider flavour branes with abelian flux).
From \eqref{localD7flch} we see that the fluxes are different on the two flavour branes. In particular, we have $\cF_0|_{\cD_{q_1}}= \frac12 H$ and $\cF_2|_{\cD_{q_1}}= \frac32 H$. This means that $\cF_0=\frac12 D_1 + \beta_0^s\cD_s +\beta_0^{q_2}\cD_{q_2} + \beta_0^b\cD_b $ and  $\cF_2=\frac32 D_1 + \beta_2^s\cD_s +\beta_2^{q_2}\cD_{q_2} + \beta_2^b\cD_b $. Again we have an ambiguity in the choice of the coefficients along $\cD_s$, $\cD_{q_2}$ and $\cD_b$.
On the other hand when we pullback $\cD_s$ and $\cD_{q_2}$ to the divisor $3D_1+\cD_s+\cD_{q_2}+\alpha^b\cD_b$ we obtain a trivial class. Then we can neglect these terms in $\cF_i$.
Making the simple choice $\beta_0^b=\beta_2^b=0$, the fluxes on the flavour branes become
\begin{equation}\label{FluxesD7Flav}
  \cF_0 = \frac{1}{2}D_1\,,
  \qquad\qquad\qquad \cF_2 = \frac32 D_1 \:.
\end{equation}

The D3-charge of the flavour branes can be determined from \eqref{ChVectExp}
after we know the class they wrap (D7-charge) and the flux living on them (D5-charge). Note however that  the actual D3-charge is given by minus the integral of the six-form.
For the two flavour branes, we have
\begin{equation}
  Q_{D3}^{D7_0^{\rm flav}} = -5\,,
  \qquad   Q_{D3}^{D7_2^{\rm flav}} = -7
   \qquad \Rightarrow\qquad  Q_{D3}^{D7_0^{\rm flav}} +Q_{D3}^{D7_2^{\rm flav}}  = -12 \:.
\end{equation}
Summarising, the charge vectors of the fractional and flavour branes are
\begin{eqnarray}
  \Gamma_{\rm frac D3} &=& 2 \Gamma_{F_0} + 3 \Gamma_{F_1} + 2 \Gamma_{F_2} =2 \cD_{q_1} +2 \cD_{q_1}\wedge D_1 - \tfrac32 d{\rm Vol}_X^0\, , \nonumber\\
 \Gamma_{D7_0^{\rm flav}}  &=&(\cD_b-\cD_{q_1}) + ( \cD_b -\cD_{q_1})\wedge \tfrac12 D_1 + 5\,
  d{\rm Vol}_X^0\, ,\\
 \Gamma_{D7_2^{\rm flav}}  &=&(\cD_b-\cD_{q_1}) + ( \cD_b-\cD_{q_1})\wedge \tfrac32 D_1 + 7\,
  d{\rm Vol}_X^0\, , \nonumber
\end{eqnarray}
where $d\textmd{Vol}_X^0$ is the normalised volume form of the CY three-fold, i.e.\ $\int_X d{\rm Vol}_X^0=1$.
Summing the three vectors gives the charge vector of the quiver
\begin{eqnarray}\label{eq:sum-charges-quiver}
  \Gamma_{\rm quiver}^{z_4=0} &=& 2 \cD_b  + 2\cD_b\wedge D_1 + \left(\tfrac{27}{2}-3\right) d{\rm Vol}_X^0\,.
\end{eqnarray}

From the above we observe that:
\begin{itemize}
\item The total D7-charge of the quiver system is the same as two D7-branes wrapping a divisor of the class of $O7_1$.
Therefore, we see that we need more D7-branes to cancel the D7-charge arising from this O7-plane.
In particular, we can simply add two branes plus their images on top of this O7-plane, realising an $SO(4)$ gauge group.
\item From \eqref{eq:sum-charges-quiver} one would na\"ively expect a globally non-vanishing  D5-charge. However by a careful analysis considering the image quiver system at $z_7=0$, one realises that the total D5-charge is cancelled.
\item The flavour branes do not introduce FW anomalies \cite{Minasian:1997mm,Freed:1999vc}: the flux has the proper (half-integral) quantisation and the wrapped cycles have no non-trivial three-cycles ($b_3=0$), so that $H_3|_{\cD_{\rm flav}}=0$.
\end{itemize}
Everything we did for the fractional D3-brane at $z_4=0$ can be done for the image at $z_7=0$. The results are exactly what one obtains by applying the orientifold involution to the charges localised at $z_4=0$.

\subsubsection*{The other D7-stacks}

As pointed out above, we need more branes than just the flavour ones to saturate the D7-tadpole. The charge vector of an O7-plane wrapping the divisor $D$ is
\begin{equation}\label{ChVectO7-pl}
\Gamma_{O7} (D)=  - 8 \, D \wedge \sqrt{\frac{\mbox{L}(\tfrac14 TD)}{\mbox{L}(\tfrac14 ND)}} =
-8 D + \frac{1}{6} D\wedge c_2(D) \:,
\end{equation}
where L$(V)=1+\frac{1}{3}(c_1(V)^2-2c_2(V))+...$ is the Hirzebruch L-genus.

Hence, we can cancel the D7-charge of the O7-planes by having two branes plus their images on top of 
$O7_1$ and four branes plus their images on top of 
$O7_2$.
We choose the B-field to be equal to $B=\frac{\cD_s}{2}$, such that $\cF_s=F_s-B=0$.
This allows us to have a non-perturbative contribution to the superpotential coming from the stack of branes on top of $O7_2$.\footnote{This condition can be relaxed when the non-perturbative effect is given by a rank-two E3-instanton~\cite{Berglund:2012gr}.}
In fact, $[O7_2]=\cD_s$ is rigid and if $\cF_s=0$, it supports a pure $SO(8)$ gauge theory which undergoes gaugino condensation.
On the other hand, the $SO(4)$ gauge group on $[O7_1]=\cD_b$ is broken to $SU(2)\times U(1)$ by the FW flux on the corresponding D7-branes. For FW anomaly cancellation we need $\cF_b+\frac{\cD_b}{2}\in H^2(\cD,\mathbb{Z})$, where we make the minimal choice $\cF_b=\frac12 D_1$.
The $U(1)$ factor decouples from the effective field theory because it becomes massive by eating up the axion given by the reduction of $C_4$ on $\cD_b$.

The D7-brane stacks on $O7_1$ and $O7_2$
have vanishing D5-charge due to the fact that the D7-branes and their images wrap the same homology class.
Regarding the D3-charge, the contribution from the two stacks is
\begin{eqnarray}
  Q_{D3}^{SU(2)}&=& 4\left(-\frac{\chi(\cD_b)}{24} -\frac12\int_{\cD_b}\cF_b^2\right) -  \frac{\chi(\cD_b)}{6}=-\frac{81}{2}\, , \nonumber\\ Q_{D3}^{SO(8)}&=&8 \left(-\frac{\chi(\cD_s)}{24}\right) -  \frac{\chi(\cD_s)}{6}=-\frac32 \, ,\nonumber
\end{eqnarray}
where the $-\chi/6$ contribution comes from the O-plane with $\chi(\cD_b)=117$ and $\chi(\cD_s)=3$.

We can now give the total D3-charge of the studied configuration:
\begin{eqnarray}
  Q_{D3}^{\rm tot} &=& Q_{D3,\,{\rm quiver}}^{z_4=0} + Q_{D3,\,{\rm quiver}}^{z_7=0} + Q_{D3}^{SU(2)} +Q_{D3}^{SO(8)}\nonumber\\
		&=& -\frac{21}{2} -\frac{21}{2} -\frac{81}{2} - \frac32 = -63\:. 
\end{eqnarray}
This negative number leaves the possibility to turn on background three-form fluxes for stabilising the dilaton and the complex structure moduli. To have a large (negative) D3-charge, one could recombine the four D7-branes on top of the O-plane wrapping $\cD_b$, to obtain a so-called Whitney-type brane \cite{Collinucci:2008pf,Collinucci:2008sq}.

In order for the whole setup to be consistent, we also need to check that the torsional K-theory charges are cancelled \cite{Witten:1998cd,Moore:1999gb}. By using the probe argument given in \cite{Uranga:2000xp}, we have checked that this is the case for the chosen B-field and brane configuration.\footnote{
One considers the set of invariant divisors in the three-fold $X$. The probe system consists of two D7-branes wrapping an invariant divisor and having zero gauge invariant flux $\cF$. Hence, the invariant divisors that do not allow a $\cF=0$ are discarded. Thereby, it can happen that the chosen B-field does not allow to cancel the possible non-zero gauge flux induced by FW anomaly cancellation. For each probe brane which wraps the remaining divisors, one needs to compute the chiral intersection with all the branes in the chosen configuration. The torsional K-theory charge is cancelled if and only if the number of $SU(2)$-fundamental is even.}

\subsubsection*{Chiral spectrum in the bulk}

Since the two flavour branes intersect each other, the image branes and the $SU(2)$ stack, the flux on the flavour branes may generate chiral matter also away from the quiver locus. Moreover, there is chiral matter on the bulk of the $SU(2)$ stack due to the flux $\cF_b$.
Let us compute the number of chiral states (see Figure~\ref{fig:setup})
\begin{eqnarray}
\# (-,+,{\bf 1}_0) &\equiv& \#\varphi_-=\langle \Gamma_{D7_0^{\rm flav}} , \Gamma_{D7_2^{\rm flav}} \rangle =   6 \:,\nonumber\\
\# (+,+,{\bf 1}_0) &\equiv& \#\varphi_+=\langle \Gamma_{D7_0^{\rm flav}}' , \Gamma_{D7_2^{\rm flav}} \rangle =   12 \:,\nonumber\\
\# (-,0,{\bf 2}_{+1}) &\equiv& \#\chi_-=\langle \Gamma_{D7_0^{\rm flav}} , \Gamma_{D7^{SU(2)}} \rangle =   0 \:,\nonumber\\
\# (+,0,{\bf 2}_{+1}) &\equiv& \#\chi_+=\langle \Gamma_{D7_0^{\rm flav}}' , \Gamma_{D7^{SU(2)}} \rangle =   9 \:,\\
\# (0,+,{\bf 2}_{-1}) &\equiv& \#\sigma_-=\langle \Gamma_{D7^{SU(2)}} , \Gamma_{D7_2^{\rm flav}} \rangle =   9 \:,\nonumber\\
\# (0,+,{\bf 2}_{+1}) &\equiv& \#\sigma_+=\langle \Gamma_{D7^{SU(2)}}' , \Gamma_{D7_2^{\rm flav}} \rangle =   18 \:,\nonumber\\
\# (0,0,{\bf 1}_{+2}) &\equiv& \#\pi=\tfrac12\left(\langle \Gamma_{D7^{SU(2)}}' , \Gamma_{D7^{SU(2)}} \rangle-\tfrac14\langle \Gamma_{O7_1} , \Gamma_{D7^{SU(2)}} \rangle\right) =   9\:,\nonumber
\end{eqnarray}
where $\Gamma_{D7^{SU(2)}}=2( \cD_b +\cD_b\wedge \tfrac12 D_1 + \frac{21}{4}d{\rm Vol}_X^0)$. The charges $\#(\pm,\pm,x_q)$ are with respect to the flavour brane $D7_0^{\rm flav}$, $D7_2^{\rm flav}$ and the $SU(2)\times U(1)$ stack. The other intersections are the images of the ones listed above.
We did not list the chiral fields at the intersection of the flavour D7-brane with its own image: in fact, their number is zero due to a cancellation occurring for the chosen wrapped divisors.
There are also chiral fields $\Phi^{\rm Adj}_{0,0}$ in the adjoint representation of $SU(2)$ whose number is counted by $h^{0,2}(\cD_b)=11$.
These scalars can be lifted by a particular class of gauge fluxes \cite{Martucci:2006ij,Bianchi:2011qh} and/or background three-form fluxes \cite{Gomis:2005wc}.

\subsubsection{Moduli stabilisation}
\label{sec:modstab}

Let us now outline how to fix both the closed and the open string moduli following the general strategy we already
described in~\cite{Cicoli:2011qg,Cicoli:2012vw}. The type IIB closed string moduli are:
\bi
\item One axio-dilaton $S=e^{-\phi}+{\rm i} C_0$;
\item $h^{1,2}_-$ complex structure moduli $U_{\alpha}$
with $\alpha=1,...,h^{1,2}_-$; \footnote{Notice that $h^{1,2}_+$ counts the number of closed string $U(1)$s.}
\item 3 orientifold even K\"ahler moduli: $T_b=\tau_b+{\rm i}\,c_{4,b}$, $T_s=\tau_s+{\rm i}\,c_{4,s}$
and $T_q=\tau_q+{\rm i}\,c_{4,q}$ where $\tau_q=\tau_{q_1} + \tau_{q_2}$, $\cD_+=\cD_{q_1} + \cD_{q_2}$, $c_{4,b}=\int_{\cD_b}C_4$, $c_{4,s}=\int_{\cD_s}C_4$ and $c_{4,q}=\int_{\cD_+}C_4$;
\item One orientifold odd K\"ahler modulus $G=b_2+{\rm i}\, c_2$ with $B_2 =b_2 \,\cD_-$ and $C_2 =c_2\,\cD_-$
where $\cD_-=\cD_{q_1} - \cD_{q_2}$.
\ei
In addition, there are open string scalars living at the quiver locus and behaving as visible sector matter fields,
and scalars living in the bulk D7 branes which support hidden sectors. These open string moduli can develop a potential either by
D-term contributions if they are charged under anomalous $U(1)$ symmetries or by F-term effects induced by supersymmetry breaking.
On the other hand, the scalar potential of the closed string moduli receives several contributions which can be classified
by taking the following expansion in inverse powers of the overall volume:
\be
V = V_D+V_F^{\rm tree} + V_F^{\rm pert} + V_F^{\rm np} \,.\nn
\ee
\begin{itemize}
\item $V_D \sim \mc{O}(1/\vo^2)$: the D-term potential includes closed string modes since fluxes on D7-branes
generate Fayet-Iliopoulos terms which depend on the K\"ahler moduli.

\item $V_F^{\rm tree} \sim \mc{O}(1/\vo^2)$: a tree-level F-term potential for the
$S$ and $U$-moduli is generated by non-trivial background fluxes
$G_3 = F_3 + {\rm i}S H_3$ which induce a tree-level superpotential $W_{\rm tree}(S,U)=\int_X G_3 \wedge \Omega$.
Given that $S$ and $U$ are fixed by imposing $D_{S,U}W=0$, 
the VEV of $V_F^{\rm tree}$ is zero \cite{Giddings:2001yu}  due to the no-scale structure at tree-level. One needs, therefore, to study subdominant perturbative and non-perturbative
corrections to $W_0 = \langle W_{\rm tree}\rangle$ and $K_{\rm tree} = - 2 \ln\vo$ in order to freeze the K\"ahler moduli.

\item $V_F^{\rm pert}\lesssim \mc{O}(1/\vo^3)$: a perturbative potential can be generated by either pure $\alpha'$ \cite{Becker:2002nn}
or $g_s$ corrections to $K$ (which appear also at different powers in $\alpha'$) \cite{Berg:2005ja,vonGersdorff:2005bf,Berg:2007wt,Cicoli:2007xp,Anguelova:2010ed,GarciaEtxebarria:2012zm,Grimm:2013gma}.

\item $V_F^{\rm np}\lesssim \mc{O}(1/\vo^3)$: non-perturbative F-term contributions can be induced by corrections to $W$
originating from E3-instantons or gaugino condensation on a D7-stack \cite{Kachru:2003aw}. Notice that non-perturbative corrections to $K$ are negligible since
we already took into account perturbative corrections to the K\"ahler potential.
\end{itemize}
Our strategy will be to stabilise the moduli order by order in a large volume expansion. Let us therefore start by considering
the D-term potential.

\subsubsection*{D-terms}

There are two anomalous $U(1)$ symmetries at the dP$_0$ singularity and three living in the bulk: one on each of the two flavour branes
and one on the stack of D7-branes on top of the $O7_1$. Therefore the total D-term potential is the sum of two contributions:
one coming from the quiver and the other from the bulk
\be
V_D = V_D^{\rm quiver} + V_D^{\rm bulk}\,.
\ee
The part from the quiver reads
\be
V_D^{\rm quiver} = \frac{1}{{\rm Re}(f_1)} \left(\sum_i q_{1i} |C_i|^2 -\xi_1 \right)^2
+ \frac{1}{{\rm Re}(f_2)} \left( \sum_i q_{2i} |C_i|^2 -\xi_2\right)^2,
\label{Dpot}
\ee
where $f_1= S + q_1 T_q$ and $f_2= S + q_2 G$ while $q_1$, $q_2$, $q_{1i}$ and $q_{2i}$
are the $U(1)$ charges of $T_q$, $G$ and the canonically normalised matter fields $C_i$, respectively.
The two Fayet-Iliopoulos terms $\xi_1$ and $\xi_2$ are given by $\xi_1 = -4 q_1\,\tau_q /\vo$
and $\xi_2 = -4 q_2\,b_2/\vo$ showing that $V_D \sim \mc{O}(1/\vo^2)$.

The potential (\ref{Dpot}) admits a supersymmetric minimum at $\xi_1=\sum_i q_{1i} |C_i|^2$
and $\xi_2=\sum_i q_{1i} |C_i|^2$. These relations fix only two moduli in terms of all the others.
In particular, there are as many flat directions as the number of open string fields charged under the anomalous $U(1)$s.
However, these flat directions can be lifted by including sub-leading F-term contributions
from supersymmetry breaking of the form $V_F \supset \sum_i m_i^2 |C_i|^2$ where $m_i \simeq M_{\rm soft}$ $\forall i$.\footnote{
We are assuming that $m_i^2>0$. In fact, if the matter fields were tachyonic, they would develop $\langle C_i \rangle \neq 0$, breaking the gauge symmetry.}
As we shall see later on, supersymmetry is broken by the bulk K\"ahler moduli which develop non-zero F-terms beyond tree-level.
This F-term potential gives a minimum at $|C_i|=0$ $\forall i$, implying $\xi_1=\xi_2=0$ or, in other words, $\tau_q = b_2= 0$,
showing that the dP$_0$ blow-up mode collapses to the singular locus.
The two anomalous $U(1)$s acquire an $\mc{O}(M_s)$ St\"uckelberg mass by eating up both the local axions, $c_{4,q}$ and $c_2$,
which therefore disappear from the low-energy theory.

The contribution to the D-term potential from the bulk looks like
\be
V_D^{\rm bulk} = \frac{1}{{\rm Re}(f_{{\rm flav},0})} D_{{\rm flav},0}^2
+ \frac{1}{{\rm Re}(f_{{\rm flav},2})} D_{{\rm flav},2}^2
+ \frac{1}{{\rm Re}(f_{D7_{O7_1}})} D_{D7_{O7_1}}^2,
\label{Dpotbulk}
\ee
where, at $\tau_{q}\rightarrow 0$, $f_{{\rm flav},0} = T_b + k_1 S$,
$f_{{\rm flav},2} = T_b + k_2 S$ and $f_{D7_{O7_1}} = T_b + k_3 S$
with $k_1$, $k_2$ and $k_3$ parameters which depend on the gauge fluxes on each stack of branes.
The three different D-terms are given by (focusing on canonically normalised matter fields)
\begin{eqnarray}
  D_{{\rm flav},0} &=& \sum_{i=1}^{12} |\varphi_+^i|^2 + \sum_{i=1}^9 |\chi_+^i|^2
  -\sum_{i=1}^{6} |\varphi_-^i|^2 - \xi_{{\rm flav},0}  \,,\nonumber\\
  D_{{\rm flav},2} &=& \sum_{i=1}^{12} |\varphi_+^i|^2 +\sum_{i=1}^{6} |\varphi_-^i|^2
  + \sum_{i=1}^{18} |\sigma_+^i|^2 + \sum_{i=1}^{9} |\sigma_-^i|^2- \xi_{{\rm flav},2}  \,,\nonumber \\
  D_{D7_{O7_1}} &=& \sum_{i=1}^9 |\chi_+^i|^2  + \sum_{i=1}^{18} |\sigma_+^i|^2
  + 2\sum_{i=1}^9 |\pi^i|^2 - \sum_{i=1}^{9} |\sigma_-^i|^2 - \xi_{D7_{O7_1}}   \,,\nonumber
\end{eqnarray}
where the Fayet-Iliopoulos terms $\xi = \frac{1}{\cV}\int_{D7} J \wedge \cF$ take the form
\be
\xi_{D7_{O7_1}}=\frac 92\,\frac{t_b}{\vo}= \left(\frac 92\right)^{2/3}\frac{1}{\vo^{2/3}}\,,\qquad \xi_{{\rm flav},0}=\xi_{D7_{O7_1}} + \frac{\sqrt{\tau_q}}{2\cV}\,,
\qquad \xi_{{\rm flav},2}=3\,\xi_{{\rm flav},0}\,.
\ee
We have seen that the D-terms from the quiver fix $\tau_q=0$.
Hence, the FI-terms of the flavour branes reduce to
$\xi_{{\rm flav},2}=3\,\xi_{{\rm flav},0}=3\,\xi_{D7_{O7_1}}=: 3\,\xi$.

The supersymmetric minimum of the bulk D-term potential (\ref{Dpotbulk}) is located at:
\bea
|\varphi_+^1|^2  &=& \sum_{i=1}^9 |\pi^i|^2-\sum_{i=1}^{9} |\sigma_-^i|^2- \sum_{i=2}^{12} |\varphi_+^i|^2+\frac{3\xi}{2}\,,  \label{D1} \\
|\varphi_-^1|^2  &=& \sum_{i=1}^9 |\chi_+^i|^2+\sum_{i=1}^9 |\pi^i|^2-\sum_{i=1}^{9} |\sigma_-^i|^2
- \sum_{i=2}^{6} |\varphi_-^i|^2+\frac{\xi}{2}\,,  \label{D2} \\
|\sigma_+^1|^2 &=& \sum_{i=1}^{9} |\sigma_-^i|^2-\sum_{i=1}^9 |\chi_+^i|^2-2\sum_{i=1}^9 |\pi^i|^2
-\sum_{i=2}^{18} |\sigma_+^i|^2+\xi\,. \label{D3}
\eea
This stabilisation procedure leaves several flat directions which will be lifted at sub-leading order by the F-term potential.
Notice that the three anomalous $U(1)$s become massive by eating up three axions given by different combinations of $c_{4,b}$
and the phases of the charged open string scalars.

\subsubsection*{F-terms}

The F-term potential receives several contributions beyond the leading order approximation:
\bi
\item Gaugino condensation on the del Pezzo divisor $\cD_s$ supporting a pure $SO(8)$ gauge theory
generates a non-perturbative superpotential of the form
\be
W_{\rm np} = A_s\,e^{-a_s\,T_s}\quad\text{with}\quad a_s=\pi/3\,. \label{Wnonpert}
\ee
Notice that the rigidity of the dP divisor guarantees the generation of non-perturbative effects.
Moreover, as explained in \cite{Cicoli:2011qg,CKM}, a `diagonal' dP divisor decouples from all the other divisors,
so avoiding any possible problem associated with the cancellation of FW anomalies
or with chiral intersections with visible sector fields.

\item The first pure $\alpha'$ correction to the action, i.e.\ corrections of order $g_s^0$ in the string coupling,
arises at order $\mc{O}(\alpha'^3)$, causing the following modification of the tree-level K\"ahler potential
\be
K = - 2 \ln \left( \vo + \frac{\zeta}{g_s^{3/2}} \right)\,,
\ee
where $\zeta = -\chi(X)\zeta(3) /[2 (2\pi)^3]\simeq 0.522$ \cite{Becker:2002nn}.

\item Corrections to the action proportional to the string coupling appear at order
$\mc{O}(\alpha'^2 g_s)$ in the open string sector \cite{Grimm:2013gma}
and $\mc{O}(\alpha'^2 g_s^2)$ in the closed string sector \cite{Berg:2005ja,vonGersdorff:2005bf,Anguelova:2010ed,GarciaEtxebarria:2012zm}.
However, at the level of the scalar potential these corrections
turn out to be negligible due to a subtle cancellation which has been called `extended no-scale structure' \cite{Berg:2007wt,Cicoli:2007xp}.
These contributions scale as $\vo^{-10/3}$ whereas the pure $\alpha'$ correction behaves as $\vo^{-3}$,
and consequently the $g_s$ effects are volume suppressed.

\item The background fluxes breaking supersymmetry generate
a gravitino mass of the order $m_{3/2} = e^{K/2} |W| \simeq W_0 /\vo$
and non-vanishing F-terms for the K\"ahler moduli in the geometric regime which look like \cite{Conlon:2006us}
\be
\frac{F^{T_b}}{\tau_b} = m_{3/2}\left[1+\mc{O}\left(\frac{1}{\vo}\right)\right]\qquad\text{and}\qquad
\frac{F^{T_s}}{\tau_s} = \frac{m_{3/2}}{\ln\left(\vo/W_0\right)}\,. \label{Fterms}
\ee
In turn, due to these non-zero F-terms, the open string scalars living in the bulk develop soft-terms which
scale as the gravitino mass suppressed by a factor of order $\ln(M_P/m_{3/2})$ \cite{Conlon:2006us}.
This F-term contribution for the fields charged under the anomalous $U(1)$ symmetries reads
(showing only the leading order expression)
\bea
V_F^{\rm matter} &\simeq&  \frac{W_0^2}{\vo^2\left[\ln\left(\vo/W_0\right)\right]^2}
\left[c_{\varphi}\left(\sum_{i=1}^{6} |\varphi_+^i|^2 + \sum_{i=1}^{12} |\varphi_-^i|^2\right)
+ c_{\chi}\sum_{i=1}^9 |\chi_+^i|^2 \right. \nn \\
&& \left. +c_{\sigma} \left(\sum_{i=1}^{18} |\sigma_+^i|^2 +\sum_{i=1}^{9} |\sigma_-^i|^2\right)+c_{\pi}\sum_{i=1}^9 |\pi^i|^2\right],
\label{Vfmatter}
\eea
where the $c$'s are $\mc{O}(1)$ coefficients which give the dependence
of the K\"ahler metric of a generic unnormalised matter field $\rho$ on the blow-up mode $\tau_s$:
$K\supset \tau_s^{c_{\rho}}\,|\rho|^2/\vo^{2/3}$. The expression (\ref{Vfmatter})
is generated by $F^{T_s}$ and it vanishes if the $c$'s are zero. In this case,
$V_F^{\rm matter}$ would be generated by the sub-leading correction to $F^{T_b}$ in (\ref{Fterms})
and it would be much more volume suppressed: $V_F^{\rm matter}\sim \vo^{-4}$~\cite{Blumenhagen:2009gk}.\footnote{The leading piece in $F^{T_b}$ cancels off with the contribution from the gravitino mass.}

\begin{figure}
\begin{center}
\includegraphics[width=0.65\textwidth]{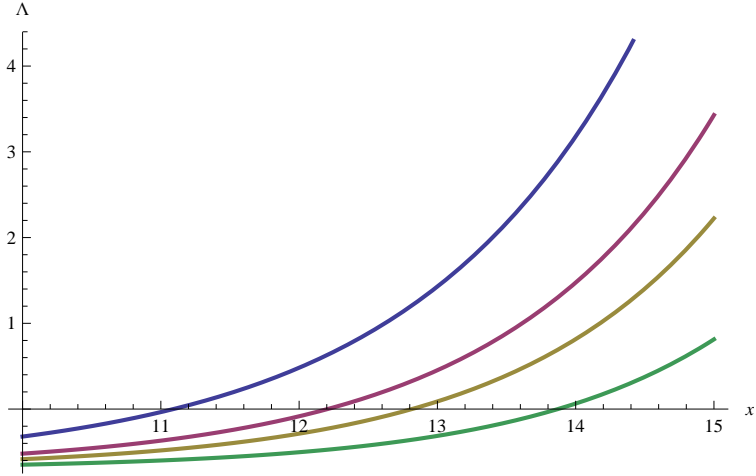}
\caption{Vacuum energy as a function of $x=\log_{10}\vo$ 
for different values of $W_0$: $W_0=1\,\text{(blue line)}$, $W_0=10^{-4}\,\text{(yellow line)}$, $W_0=10^{-7}\,\text{(purple line)}$, $W_0=10^{-14}\,\text{(green line)}$;} \label{Lambda}
\end{center}
\end{figure}

Substituting the stabilisation obtained by imposing $V_D^{\rm bulk}=0$, we end up with
\bea
V_F^{\rm matter} &\simeq&  \frac{W_0^2}{\vo^2\left[\ln\left(\vo/W_0\right)\right]^2}
\left[\left(c_{\varphi} +c_{\chi}- c_{\sigma}\right)\sum_{i=1}^9 |\chi_+^i|^2
+ 2\left(c_{\sigma}-c_{\varphi} \right) \sum_{i=1}^{9} |\sigma_-^i|^2 \right. \nn \\
&& \left. + \left(2 c_{\varphi} +c_{\pi} - 2 c_{\sigma}\right) \sum_{i=1}^9 |\pi^i|^2
+ \left(2 c_{\varphi}  + c_{\sigma}\right) \xi\right]. \label{Fmatter}
\eea
If $c_\chi>c_\sigma-c_\varphi>0$ and $c_\pi>2\left(c_\sigma-c_\varphi\right)>0$, the potential (\ref{Fmatter})
has a minimum at $\chi_+^i=\sigma_-^i=\pi^i=0$ $\forall i$, simplifying the previous expression to
\be
V_F^{\rm matter} \simeq p\,\frac{W_0^2}{\vo^{8/3}\left[\ln\left(\vo/W_0\right)\right]^2}\,,
\quad\text{with}\quad p:= \left(2 c_{\varphi}  + c_{\sigma}\right)\left(\frac 92\right)^{2/3}\,,
\ee
whereas the minimisation equations (\ref{D1}) to (\ref{D3}) become
\bea
|\varphi_+^1|^2  &=& - \sum_{i=2}^{12} |\varphi_+^i|^2+\frac{3\xi}{2}\,,  \label{d1} \\
|\varphi_-^1|^2  &=& - \sum_{i=2}^{6} |\varphi_-^i|^2+\frac{\xi}{2}\,,  \label{d2} \\
|\sigma_+^1|^2 &=& -\sum_{i=2}^{18} |\sigma_+^i|^2+\xi\,. \label{d3}
\eea
We will now show how to stabilise the overall volume of the CY, fixing the value of $\xi$.
However, several open string directions are still flat. Their total number is:
11 from (\ref{d1}) + 5 from (\ref{d2}) + 17 from (\ref{d3}) = 33
on top of all the phases of the open string fields. In order to lift these flat directions,
one has to include more F-term contributions for matter scalars beyond the na\"ive supersymmetry
breaking effects that we took into account.
\ei
Summing up all these different effects, the total F-term potential takes the form
\be
V_F \simeq \frac 83  (a_s A_s)^2  \sqrt{\tau_s}\, \frac{e^{-2\, a_s \tau_s}}{\vo}
   - 4 \,a_s A_s W_0 \tau_s\frac{e^{-a_s \tau_s}}{\vo^2} + \frac{3}{4} \frac{\zeta W_0^2}{g_s^{3/2} \vo^3}
   +p\,\frac{W_0^2}{\vo^{8/3}\left[\ln\left(\vo/W_0\right)\right]^2}\,.
\label{VFtotal}
\ee
In the regime $a_s\tau_s\gg 1$, the minimisation with respect to $\tau_s$ gives
\be
e^{-a_s \tau_s}= \frac{3\sqrt{\tau_s}}{4 a_s A_s}\frac{W_0}{\vo}
\qquad\Rightarrow\qquad a_s \tau_s \simeq\ln\left(\vo/W_0\right)\,.
\label{tsVEV}
\ee
Plugging this result into (\ref{VFtotal}), we find
\be
V = \frac{W_0^2}{\vo^3} \left\{\frac{3\zeta}{4 g_s^{3/2}}
-\frac 32\left[\frac{\ln\left(\vo/W_0\right)}{a_s}\right]^{3/2}
   + p\,\frac{\vo^{1/3}}{\left[\ln\left(\vo/W_0\right)\right]^2} \right\}.
\label{VFfinal}
\ee
Minimising with respect to $\vo$ we obtain
\be
\frac{3\zeta }{4g_s^{3/2}}=\frac 32 \left[\frac{\ln\left(\vo/W_0\right)}{a_s}\right]^{3/2}
\left(1-\frac{1}{2\ln\left(\vo/W_0\right)}\right)-\frac 89
\,p\,\frac{\vo^{1/3}}{\left[\ln\left(\vo/W_0\right)\right]^2}\left(1+\frac{3}{4\ln\left(\vo/W_0\right)}\right)\,,
\label{tbVEV}
\ee
which substituted in (\ref{VFfinal}) yields the following expression for the vacuum energy
\be
\Lambda\equiv\langle V \rangle = \frac{W_0^2}{\langle\vo\rangle^3} \sqrt{\ln\left(\frac{\langle\vo\rangle}{W_0}\right)}\left\{-\frac{3}{4\,a_s^{3/2}}
+ \frac{p}{9}\frac{\langle\vo\rangle^{1/3}}{\left[\ln\left(\vo/W_0\right)\right]^{5/2}}\left(1-\frac{6}{\ln\left(\vo/W_0\right)}\right) \right\}.
\label{Vmin}
\ee
Setting $a_s=\pi/3$ and writing $\vo = 10^x$, Figure~\ref{Lambda}
shows how the
vacuum energy changes as a function of $x$ for different values of $W_0$ at constant $p$ (shown here for $c_\sigma=1$ and $c_\phi=1/2$).

The preferred values of $\vo$ and $W_0$ are chosen in such a way to obtain a Minkowski vacuum
and TeV-scale supersymmetry at the same time. In the presence of flavour branes,
loop corrections to the visible sector gauge kinetic function might induce moduli redefinitions of
the form $\tau_q \to \tau_q + \alpha \ln\vo$ which can de-sequester the visible sector~\cite{Conlon:2010ji,Choi:2010gm}.
Thus, the soft terms generated by gravity mediation scale as \footnote{Note that the presence of these field redefinitions is still under active discussion, see for example~\cite{deAlwis:2012bm}. If they are absent, the visible sector is completely sequestered, resulting in soft masses
of the order $M_{\rm soft}\sim M_P /\vo^{3/2}$ or even smaller~\cite{Blumenhagen:2009gk}.}
\be
M_{\rm soft} \simeq \frac{m_{3/2}}{\ln\left(M_P/m_{3/2}\right)} \simeq \frac{W_0\,M_P}{\vo \ln\left(\vo/W_0\right)}\,.
\ee
Requiring these soft terms to be around the TeV-scale, the ratio $\vo/W_0$ is constrained to be
of order $\vo/W_0 \simeq 5\cdot 10^{13}$. From (\ref{Vmin}) one can find numerically that this
is satisfied with also a Minkowski solution for $W_0\simeq 0.01$ and $\vo\simeq 5\cdot 10^{11}$.
Plugging these numbers in (\ref{tbVEV}), we find that for $\zeta \simeq 0.522$, the string coupling
has to be $g_s\simeq 0.015\simeq 1/65$, i.e.\ in the weak coupling regime.\footnote{Notice that these results slightly 
depend on the value of the parameter $p$ which is however expected to be of order unity. In fact, 
the results quoted in the main text are obtained for $p\simeq 5.45$. If we change this value to $p\simeq 1$ by 
considering different values for $c_\sigma$ and $c_\phi$, we would obtain $W_0\simeq 1$ and $\vo \simeq 5\cdot 10^{13}$ 
but the same value of $g_s$.} The string scale turns out to be intermediate $M_s \sim M_P/\sqrt{\vo}\sim 10^{12}$ GeV and corresponds to the unification scale for the left-right symmetric model under examination, as obtained
in \cite{Aldazabal:2000sa,Aldazabal:2000sk,Ibanez:2012zz}. Notice that if we substitute the requirement of a vanishing cosmological constant
with the one of getting the right unification scale, then this last phenomenological constraint would imply $\Lambda\simeq 0$.

\subsection{Phenomenology of the left-right model}

The left-right model has several interesting phenomenological features. It contains an observable sector with gauge symmetry $SU(3)_c\times SU(2)_L\times SU(2)_R\times U(1)_{\textmd{B-L}}$ with three families of quarks, leptons and Higgses. As shown in Figure~\ref{fig:dp0quiver}, there are additional $SU(3)_c$ exotics, denoted by $A$ and $\tilde{A}$ which however do allow for the interesting coupling with the D7-D7 state $\varphi_-$
\begin{equation}
W\supset A\varphi_-\tilde{A}\, .
\end{equation}
As shown in~\eqref{d2}, $\varphi_-$ can obtain a non-vanishing VEV and this VEV can generate a mass as high as the string scale for the $SU(3)_c$ exotics. One may also directly recombine the two flavour D7-branes $D7^{\rm flav}_0$ and $D7^{\rm flav}_2$ into a flavour brane of type $D7^{\rm flav}_1$, i.e.\ relative to the node $m_1$ in Figure~\ref{fig:dp0quiverflav2}; this would wrap one connected representative of the homology class $2\cD_b-2\cD_{q_1}$ and have the flux $\cF_{\rm flav}^1=D_1$ such that the local flavour branes is the right one.\footnote{This different configuration would not change drastically the scales obtained after moduli stabilisation.}
Once we have realised the decoupling of the Standard Model exotics, we are only left with the Standard Model matter content with three families of Higgses and right handed neutrinos, which we shall assume in the rest of this section.

Given this matter content, the superpotential of the D3-D3 and D3-D7 states is given by
\begin{equation}
W_{\rm matter}=y_{ijk}Q_L^i H^j Q_R^k=(\lambda_{\rm local} \epsilon_{ijk}+\lambda_1 |\epsilon_{ijk}|+\lambda_2\delta_{ijk})Q_L^i H^j Q_R^k\,
\label{eq:superpotlocal}
\end{equation}
The Yukawa coupling proportional to $\lambda_{\rm local}\epsilon_{ijk}$ is the coupling appearing in a non-compact dP$_0$ model without any geometric deformations. The terms proportional to $\lambda_{2,3}$ capture the possible deformations that can arise from non-commutative deformations of the background~\cite{Wijnholt:2005mp} or 
by taking into account compactification effects~\cite{Maharana:2011wx,Burgess:2011zv}. As pointed out in~\cite{Conlon:2008wa}, the local Yukawa coupling leads to the phenomenologically undesirable mass hierarchy of type $(0,M,M)$ with two degenerate mass eigenvalues. By including corrections proportional to $\lambda_{1,2}$ this mass structure can be changed. For example, taking the limit $\lambda_{\rm local}\approx \lambda_1 \ll \lambda_2,$ the Yukawa couplings become diagonal to leading order, resulting in three distinct mass eigenvalues proportional to $m_{\rm quark}^2\simeq\lambda_2^2(|H_1|^2,|H_2|^2,|H_3|^2).$\footnote{This diagonal structure of the Yukawa coupling might be very advantageous to prevent large flavour changing neutral currents as it allows no `off-diagonal' couplings for the Higgses `orthogonal' to the Standard Model Higgs. For a detailed recent discussion on how to evade flavour changing neutral currents and a discussion of experimental limits see~\cite{Gupta:2009wn}.} The inclusion of corrections proportional to $\lambda_{1,2}$, in principle, opens the possibility to generate hierarchical masses even in dP$_0.$ However, the Yukawa couplings for up- and down-quarks in this left-right model are equal at tree-level and hence a desired flavour mixing to reproduce the hierarchical structure in the CKM matrix as discussed in~\cite{ Krippendorf:2010hj,Dolan:2011qu} is not possible. To evade this constraint, one can for example consider realisations of this model on singularities coming from higher del Pezzo surfaces. In addition, note that the above superpotential~\eqref{eq:superpotlocal} does not contain any Yukawa couplings for the leptons and no $\mu$-term.

\begin{figure}
\begin{center}
 \includegraphics[width=0.7\textwidth]{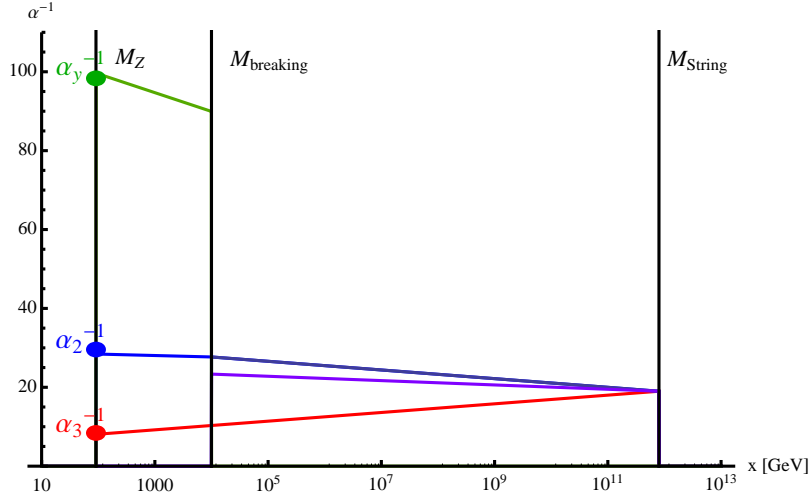}
 \end{center}
\caption{Gauge coupling unification in the left-right symmetric model for a unified coupling $\alpha_{\rm unif}^{-1}=19,$ a string scale $M_{s}=9\cdot 10^{11} {\rm GeV},$ and a breaking scale of $SU(2)_R\times U(1)_{\textmd{B-L}}\to U(1)_Y$ at $10\ {\rm TeV}$ after which we assume only one pair of Higgses to be massless. The running of the various couplings is colour-coded as follows: $\alpha_3^{-1}$ (red), $\frac{3}{32} \alpha^{-1}_{\textmd{B-L}}$ (purple), $\alpha^{-1}_{2L,2R}$ (dark-blue), and $\alpha_{Y}^{-1}$ (green). The experimentally observed values of the gauge couplings at $M_Z$ are indicated with the respective disks.}\label{fig:unification}
\end{figure}

Due to a non-standard normalisation of $U(1)_{\textmd{B-L}}$ of $k=32/3,$ the tree-level Weinberg angle is given by $\sin\theta_W=0.214$ which is close to the experimentally observed value. This Weinberg-angle and the presence of three generations of Higgses  lead to gauge coupling unification at a similar level as the standard GUT scale MSSM, but at a unification scale of order $M_s\simeq 10^{12}\,\textmd{GeV}$  \cite{Aldazabal:2000sa,Aldazabal:2000sk,Ibanez:2012zz}. To achieve unification, one assumes a breakdown of $SU(2)_R\times U(1)_{\textmd{B-L}}\to U(1)_Y$ near the weak scale which represents the natural scale for this breakdown as it needs to occur radiatively and, hence, is tied to the breakdown of supersymmetry. The unified coupling is given by (see~\cite{Malyshev:2007zz})
\begin{equation}
  \alpha_{\rm unif}^{-1} \equiv \alpha_0^{-1}  =\alpha_1^{-1}  =\alpha_2^{-1} =  {\rm Re}(S) |Z_{\rm frac}| = g_s^{-1}/3 \:,
\end{equation}
where $Z_{\rm frac}=1/3$ is the central charge of the fractional branes (when the dP$_0$ is shrunk to zero size) \cite{Douglas:2000gi}.
It is interesting to notice that the value $g_s^{-1}\simeq 65$, necessary for TeV-scale soft terms, gives exactly the correct phenomenological value $\alpha^{-1}_{\rm unif} \simeq 20$.
The evolution of gauge couplings is shown in Figure~\ref{fig:unification}.

As mentioned in the introduction, this `coincidence' is highly non-trivial since two `parameters'\footnote{We are using the fact that fluxes provide a discretuum of values of $W_0$ and $g_s$ and use these quantities as parameters. Notice that the number of flux vacua depends exponentially on the value of $h^{1,2}$ which is large enough in this case $h^{1,2}=112$. Also at this stage the cosmological  constant value needs only to be cancelled up to  the supersymmetry breaking scale since as usual there will be quantum contributions to the vacuum energy at lower energies for which we are assuming the landscape approach to the cosmological constant problem~\cite{Bousso:2000xa}. In order for the tuning to be efficient just a small (positive) value is needed at this stage. Quantum corrections to the cosmological constant below the supersymmetry breaking scale can be cancelled by a further tuning of $W_0$.},  $g_s$ and $W_0$ (determined by the fluxes via dilaton and complex structure moduli stabilisation as in~\cite{Giddings:2001yu}), together with the value of the volume at the minimum of the scalar potential (which is also determined as a function of $W_0$ and $g_s$) are enough to determine four physical quantities: the string or unification scale, $\alpha^{-1}_{\rm unif} $, the cosmological constant and the scale of soft terms. The values of these physical quantities agree with the experimental data with in addition the prediction of soft terms around the TeV scale. 
This addresses the hierarchy problem and leads to a possible contact with the LHC experiment. 
The fact that both $\alpha^{-1}_{\rm unif} $ and the unification energy scale obtained in this way precisely agree with the low energy calculations based on the low-energy spectrum and RG running of the couplings to high energies is remarkable. It may turn out to be only a happy coincidence in this case but  at the very least illustrates the challenge that general string models will have to face when they reach the level of addressing gauge coupling unification after moduli stabilisation.

\section{Conclusions and outlook}
\label{sec:conclusions}

In this article we successfully extended our previous construction of global models with D3-branes at singularities and moduli stabilisation \cite{Cicoli:2012vw}, to the class of models including both D7 flavour and D3-branes at singularities with moduli stabilisation.

We restricted our attention to local models with D3/D7-branes where all the flavour branes are connected objects also in the resolved picture. Although this class of models does not allow for a realisation of many local models,
it still provides a very rich structure.

In this paper we concentrated on models arising from the dP$_0$ singularity just for simplicity and to be as explicit as possible. Within this context we managed to provide a global embedding to the left-right symmetric model at dP$_0$. This model is such that the low energy spectrum together with the $U(1)$ normalisation gives rise to gauge coupling unification at an intermediate scale. We found that after embedding the model in a compact CY orientifold with de Sitter moduli stabilisation and TeV-scale soft masses, both the scale of unification and the value of the unified coupling can be dynamically determined to agree with the low-energy calculations running backwards the RG equations to high energies from the low-energy spectrum.
It would be interesting to better understand the general conditions required for this coincidence to happen for more generic models.

Even though this model has very promising features, there are challenges regarding the structure of Yukawa couplings. Although the inclusion of non-commutative deformations of the geometry or bulk effects can lift the degenerate mass eigenvalues $(0,M,M)$ and lead to hierarchical quark masses, the hierarchical flavour mixing parametrised in the CKM matrix cannot be obtained due to the unification of up- and down-type Yukawa couplings. Furthermore, the absence of lepton Yukawa couplings and the $\mu$-term represent additional phenomenological challenges. To avoid these problems a promising avenue is to extend this construction to higher del Pezzo singularities that have been shown to lead to a more realistic Yukawa structure, allowing for hierarchical masses and mixing as in the CKM and PMNS mixing matrices without relying on bulk effects. We will leave the construction of a global completion of such models 
for a future publication.
Phenomenologically there are additional important constraints that have to be considered. Probably the most serious are to give masses to the extra Higgs fields to keep consistency with FCNC constraints, to explicitly realise the radiative breakdown to the Standard Model gauge group, and
to solve the cosmological moduli problem associated with the light volume mode with mass of the order $1$~MeV.

In summary, in this paper we have made substantial progress towards fully controlled globally embedded local models with stabilised moduli. However, there are many remaining open questions to be explored in order to achieve a full understanding of this general class of models and extract truly realistic properties regarding phenomenological and cosmological questions. We hope to return to these remaining open questions in the near future.

We finally point out an interesting observation.
From the study of the charge vectors of D3- and D7-branes performed in section~\ref{BraneSetUp}, we observe that the total charge of the quiver system
under consideration might be realised by another configuration, i.e.~three D3-branes at the singularity realising an $SU(3)^3$ gauge group, plus two D7-branes wrapping the divisor $\cD_b$ with fluxes respectively equal to zero and $\cF=\frac12D_1,\,\frac32D_1$.
This suggests a possible smooth transition between the two systems.
In this sense, the local quiver theories which can be consistently embedded globally, might be continuously connected
to each other by supersymmetric transitions involving D7-branes coming from the bulk.
We leave the detailed study of these transitions for a forthcoming publication~\cite{Cicoli:2013zha}.

\acknowledgments
It is a pleasure to thank Andres Collinucci, I\~naki Garc\'ia-Etxebarria, Noppadol Mekareeya and Angel Uranga for discussions. The work of CM and SK was supported by the DFG under TR33 ``The Dark Universe''. SK was also supported by the European Union 7th network program Unification in the LHC era (PITN-GA-2009-237920).
SK and  CM thank ICTP for hospitality. We thank the referee for asking relevant questions that prompted modification of Section \ref{modifsect} of the original version.

\appendix
\section{Disconnected flavour branes on the resolution of \texorpdfstring{$\mathbb{C}^3/\mathbb{Z}_3$}{C3/Z3}}\label{appendix}

Consider the non-compact space $\mathbb{C}^3/\mathbb{Z}_3$. A flavour D7-brane brane is a brane which passes through the orbifold singularity \cite{Franco:2006es}. 
From string theory, we know that for any four-cycle in the quotient space there are three different discrete choices that define a D7-brane passing through the singularity. These three different choices are the three kinds of flavour branes 
(associated with the nodes $m_0$, $m_1$, $m_2$ in the quiver diagram in Figure \ref{fig:dp0quiverflav2}). In $\mathbb{C}^3/\mathbb{Z}_3$ all the three consistent flavour branes are holomorphic objects.
 
Now, let us go to the resolution of this space, i.e. the total space of $\cO_{\mathbb{P}^2}(-3)$, that is the canonical bundle of $\mathbb{P}^2$. This space $X$ is a non-compact toric Calabi-Yau,  whose weight matrix is
\begin{equation}
\begin{array}{|c|c|c|c|c|c|c|c||c|}
\hline z_1 & z_2 & z_3 & y  \tabularnewline \hline \hline
    1  &  1  &  1  & -3  \tabularnewline\hline
\end{array}
\end{equation}
The locus $y=0$ is the exceptional divisor, i.e. the $\mathbb{P}^2$ parametrised by the homogeneous coordinates $z_i$. We have the following linear relations between the four divisors: $D_{z_1}=D_{z_2}=D_{z_3}\equiv D_H$ and $D_y=-3D_H$. The restriction of $D_H$ on the blown up dP$_0$ is equal to the hyperplane class $H$.

As we have seen, the local charges of the flavour branes are determined by the number of chiral states they have at the intersection with the fractional branes. Hence, given a consistent set of flavour branes described by a quiver, we know what are the local charge vectors of these flavour branes in the resolved picture. We have seen that some flavour branes can have negative local D7-charge. These branes are consistent supersymmetric branes in $\mathbb{C}^3/\mathbb{Z}_3$. 

Let us consider the example $m_0=m_1=m_2=1$. From equation \eqref{GenericFlavChVct}, we see that the $m_0$- and the $m_2$-flavour branes have local charge $+H$. They have to wrap a four-cycle in the homology class $D_H$. This class has connected components given by the vanishing of a polynomial of degree 1 in the $z_i$. The $m_1$-flavour brane has local charge $-2H$ and should then wrap the class $-2D_H$; the most generic equation describing a four-cycle in this class is
\begin{equation}
y \cdot P_1(z_i) = 0 
\end{equation}
where $P_1$ is a polynomial of degree 1 in the $z_i$. We see that this brane is actually reducible in two pieces: one of which is at $\{y=0\}$, i.e.\ the exceptional $\mathbb{P}^2$.

Hence, the flavour branes with negative local D7-charge are described by disconnected pieces in the large volume limit of the moduli space. At the orbifold point they should however form a bound state due to some $\alpha'$ effect.

\bibliographystyle{JHEP}
\bibliography{CKMQV_2}

\end{document}